\documentclass[aps,pra,showkeys,twocolumn,superscriptaddress]{revtex4-2}
\usepackage[utf8]{inputenc}
\usepackage{amsmath,graphicx,array,amsthm}

\usepackage[colorlinks={true}]{hyperref}
\hypersetup{colorlinks=true,linkcolor=red,citecolor=magenta,urlcolor=blue}
\usepackage{subfigure}
\usepackage{indentfirst}
\usepackage{braket}
\usepackage{diagbox}
\usepackage{amssymb}
\usepackage{bbold}
\usepackage{colortbl}
\usepackage{multirow}
\usepackage{orcidlink}
\usepackage{pstricks}

\begin{document}

\title{Quantumness of gravitational cat states in correlated dephasing channels}

\author{Saeed Haddadi \orcidlink{0000-0002-1596-0763}}\email{haddadi@semnan.ac.ir}
\affiliation{Faculty of Physics, Semnan University, P.O. Box 35195-363, Semnan, Iran}

\author{Mehrdad Ghominejad \orcidlink{0000-0002-0136-7838}}\email{mghominejad@semnan.ac.ir}
\affiliation{Faculty of Physics, Semnan University, P.O. Box 35195-363, Semnan, Iran}

\author{Artur Czerwinski \orcidlink{0000-0003-0625-8339}}
\email{aczerwin@umk.pl}
\affiliation{Institute of Physics, Faculty of Physics, Astronomy and Informatics, Nicolaus Copernicus University in Torun, ul. Grudziadzka 5, 87–100 Torun, Poland}
\affiliation{Single Photon Applications Laboratory, National Laboratory of Atomic, Molecular and Optical Physics, Torun, Poland}

\date{\today}

\begin{abstract}
We study the quantumness of gravitational cat states in correlated dephasing channels. Our focus is on exploring how classical correlations between successive actions of a dephasing channel influence the decoherence of two gravitational cats (two qubits) at a thermal regime. The results show that the quantum coherence, local quantum Fisher information, and Bell non-locality can be significantly enhanced by augmenting classical correlations throughout the entire duration when the two qubits pass the channel. However, the gravitational interaction and energy gap between states exhibit intricate impacts on the quantum characteristics of gravitational cats. New features are reported that can be significant for both gravitational physics and quantum information processing.
\end{abstract}

\maketitle
\vspace{-1.25cm}

\section{Introduction}
Gravitational cat (gravcat) states represent an intriguing concept at the intersection of quantum mechanics and gravitational physics \cite{Anastopoulos2015}, which gives rise to the interesting notion of macroscopic superpositions in gravitational fields \cite{zhengprl2023}. Derived from the foundational principles of quantum mechanics, where particles can exist in multiple states simultaneously, the concept of ``cat states" comes from the famous thought experiment of Schrödinger's cat, in which a hypothetical cat can be both alive and dead until observed. Extending this concept to gravitational fields presents profound implications for our understanding of quantum gravity and the behavior of matter at the macroscopic scale \cite{Abdo2009,Hossenfelder2013,Modesto2017}.
In gravcat states, the gravitational field acts as an environment in which macroscopic objects, similar to Schrödinger's cat, exist in superpositions of distinct gravitational states. These states are not only of theoretical interest but also hold promise for experimental exploration, suggesting a unique window into the interplay between gravity and quantum mechanics \cite{
Anastopoulos2020,Anastopoulos2021}.

Historically speaking, there has been ongoing discussion around a fundamental question related to this issue: how can we confirm if gravity must be treated as a quantized phenomenon or ``why we need to quantise everything, including gravity" \cite{Marlettonpj2017}?  Furthermore, is there a universally applicable experimental method that can determine whether gravity operates on a quantum level \cite{MarlettoPRL2017,Bose2017}? According to certain perspectives drawing from quantum information theory, it is posited that an interaction capable of generating entanglement between two systems must inherently possess quantum characteristics. Consequently, a significant indicator of quantum gravity would be the observation of entanglement between massive states brought about by gravitational interactions \cite{Rovelli2021,BosePRD2022}. Contrary to current tests that rely on detecting entanglement mediated by gravity, Lami et al. \cite{Lami2024PRX} have recently proposed a new approach focusing solely on coherent states. Interestingly, their method eliminates the need to generate extensively delocalized states of motion or detect entanglement, which does not occur at any stage of the process.

Hence, the study of gravcat states has attracted considerable attention in recent years \cite{Dahbi2022,Rojas2023,Atta2023,Saeed2024}, which has been stimulated by advances in theoretical frameworks and experimental techniques. Some researchers have proposed various schemes to generate and observe gravcat states, utilizing tools from gravitational wave detection, quantum optics,  and precision measurement techniques. These efforts not only deepen our knowledge of fundamental physics but also open avenues to explore the quantum nature of gravity in new ways. Some papers \cite{ref1,ref2,ref3,ref4} highlight theoretical proposals and experimental approaches towards understanding and realizing gravcat states.

The quantum coherence of gravcat states is influenced by various factors, including environmental noise, gravitational interactions, and generally decoherence mechanisms. Environmental noise, arising from factors such as thermal fluctuations or stray gravitational fields, can disrupt the delicate quantum superpositions required for coherence. In particular,  gravitational interactions pose notable challenges because gravity is a long-range force that can lead to uncontrollable interactions between massive objects and their surroundings. Understanding the quantum coherence of gravcat states is crucial for probing the boundaries between quantum mechanics and gravity and clarifying the nature of space-time at macroscopic scales.

Moreover, correlated dephasing channels introduce noise into our quantum system and cause the loss of coherence between the different components of the superposition \cite{Palma2002}. This leads to the degradation of the gravcat state, eventually collapsing it into a classical mixture of states. Therefore, correlated dephasing channels can disrupt quantum coherence and quantum correlations between the constituent parts of the gravcat states.

When considering the quantumness of gravcat states in correlated dephasing channels, it is usually possible to analyze how these states evolve under the influence of decoherence. As we know, decoherence processes due to gravitational interactions with the environment can lead to the suppression of quantum superpositions and the emergence of classical behavior. Hence, we are motivated to scrutinise some parameters such as the strength of the gravitational interaction, the energy gap between the ground and excited states, the nature of the environment, and the degree of correlation between different components of the system to understand how these factors affect the perseverance of quantum features in gravcat states.

In this study, we raise a fundamental query lacking in the existing literature: what is the impact of correlated dephasing channels on the quantum characteristics of gravcat states? While recent research has delved into the influence of temperature on the quantum coherence and quantum correlations of gravcat states \cite{Dahbi2022,Rojas2023,Atta2023,Saeed2024} (and other states \cite{DW01,DW02,DW03,DW04,EPJChaddadi2024}) within closed and open systems, our focus is on exploring how correlated dephasing channels influence the quantumness of these states. Strictly speaking, we investigate the decoherence dynamics of two qubits undergoing a classically correlated dephasing channel \cite{Palma2002}. Specifically, when we refer to a channel as classically correlated, we imply the existence of classical correlations between its successive operations on the two qubits. Using this framework, we will demonstrate that these classical correlations effectively delay the decay of quantum correlation and coherence.

\begin{figure}[t]
  \centering
  \includegraphics[width=3in]{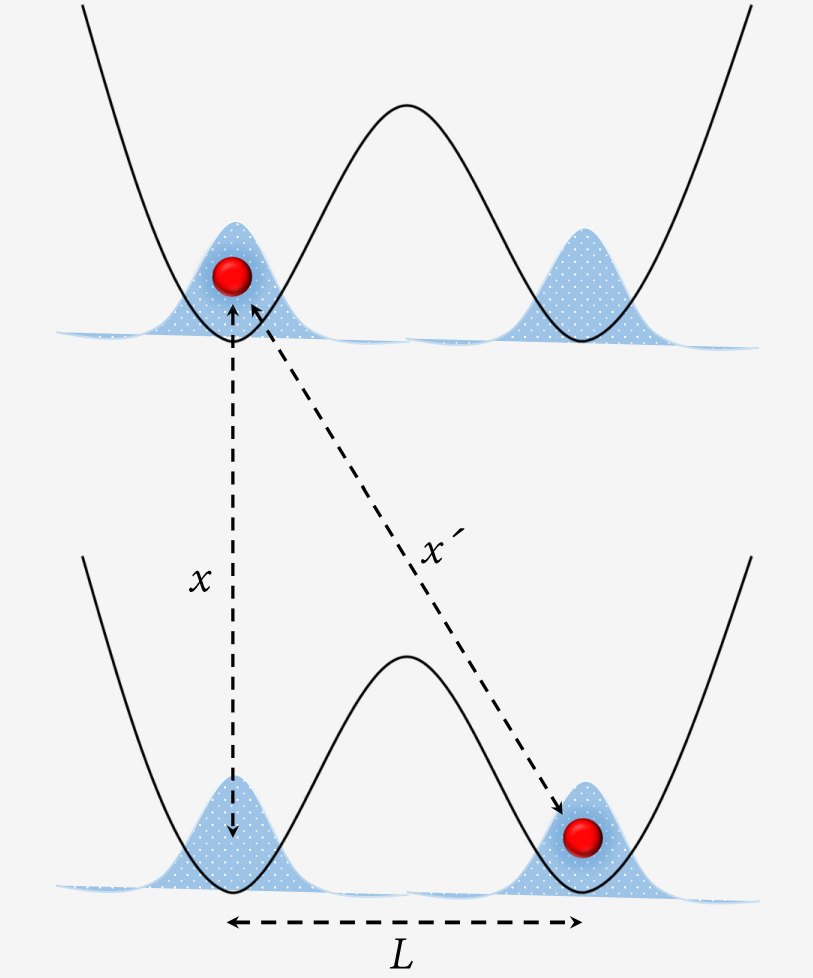}
  \caption{Schematic of the gravcats model's geometry involves symmetric double-well potentials positioned along separate axes. These axes are parallel to each other, with a gap between them represented by the distance $x =\sqrt{{x^{\prime}}^2-L^2}$.}\label{figure1}
\end{figure}

\section{Gravcat model and method}\label{sec2}
We consider here a simple model for the interaction of two gravcats that each of them corresponds to a qubit. The Hamiltonian describing this physical system would be presented as \cite{Anastopoulos2020}
\begin{equation}\label{eq2}
H=H_{\omega}+H_{\gamma},
\end{equation}
with $H_{\omega}=\omega(\sigma_z \otimes \mathbf{I}_2 + \mathbf{I}_2 \otimes \sigma_z)/2$ and $H_{\gamma}=-\gamma (\sigma_x \otimes \sigma_x)$, where $\sigma_{x}$ and $\sigma_{z}$ are Pauli operators and $\mathbf{I}_2$ is identity operator. Additionally,  $\omega$ represents the energy gap between the ground state and the first excited state, and  $\gamma=\frac{G m^ 2}{2} \left(\frac{1}{x} - \frac{1}{x^{\prime}}\right)$ serves as a parameter controlling the strength of the gravitational interaction between states with masses $m$. Herein, $G$ denotes the universal gravitational constant, and $x$ ($x^{\prime}$) represents the distance between two masses when they are at the same (different) relative minimum (look at Fig. \ref{figure1}).

In the context of statistical mechanics, the Gibbs density operator, denoted by $\rho$, represents the density matrix that describes the statistical state of a system at thermal equilibrium. Mathematically, for a system with a Hamiltonian $H$ at absolute temperature $T$, the Gibbs density operator is given by
\begin{equation}\label{eq3}
\rho_T=\frac{e^{-H/k_{B}T}}{Z},
\end{equation}
where $Z$, the partition function, is a normalization factor given by $Z=\textmd{tr}[\exp(-H/k_{B}T)]$ with the Boltzmann constant $k_{B}$ (throughout this paper $k_{B}=1$ is considered). The operator \eqref{eq3} provides a way to compute the expectation values of observables in the system, and satisfies the following conditions: $\rho=\rho^{\dag}$, $\textmd{tr}\rho=1$, and $\rho\geq0$.

Due to the functional equation expressed in \eqref{eq3} and the structure of our Hamiltonian as given in  Eq. \eqref{eq2}, the Gibbs density matrix would be obtained as \cite{Rojas2023,Atta2023,Saeed2024}
\begin{equation}\label{eq4}
\rho_T=\left(\begin{array}{cccc}
a^- & 0 & 0 & c\\
0 & b & d & 0\\
0 & d & b & 0\\
c & 0 & 0 & a^+
\end{array}\right),
\end{equation}
where the nonzero entries read
\begin{align}\label{eq5}
a^\pm= & \frac{\alpha\cosh \left(\alpha/T\right)\pm\omega  \sinh \left(\alpha/T\right)}{Z\alpha}, \quad b=\frac{\cosh \left(\gamma /T\right)}{Z},\nonumber \\
c=& \frac{\gamma  \sinh \left(\alpha/T\right)}{Z \alpha},\quad d=\frac{\sinh \left(\gamma/ T\right)}{Z},
\end{align}
with $Z=2 \left[\cosh \left(\alpha/T\right)+\cosh \left(\gamma /T\right)\right]$ and $\alpha^{2}=\omega ^2+\gamma ^2$.

\section{Gravcats in correlated dephasing channel}\label{sec3}
In this section, we review the initial details concerning the characterization of the correlated quantum channel. To simplify matters, we focus on a two-qubit system initially prepared in state $\rho(0)$. Let's imagine these two qubits sequentially pass through a channel $\mathcal{E}$. If the channel operates independently and uniformly on each qubit, the resulting state of the system can be represented by the map \cite{Nielsen2000}
\begin{equation}\label{Kraus1}
\rho(t)=\mathcal{E}[\rho(0)]=\sum_{i, j=0}^3 E_{i j} \rho(0) E_{i j}^{\dagger},
\end{equation}
where $E_{i j}=\sqrt{p_i p_j} \sigma_i \otimes \sigma_j$ are the Kraus operators with identity operator $\sigma_0=\mathbf{I}_2$ and three Pauli operators $\sigma_{1, 2, 3}= \sigma_{x, y, z}$. The probability distribution is defined by $\{p_i\}$  with $\sum_i p_i = 1$ and $p_i \geq 0$. Besides, $\{E_{i j}\}$ satisfy the completely positive and trace preserving condition as $\sum_{ij} E_{i j}^{\dagger} E_{i j}=\mathbf{I}_4$.

The map depicted above characterizes the scenario where the channel $\mathcal{E}$ operates without memory, meaning its impact on qubit $A$ does not influence its behavior towards the subsequent qubit $B$. However, this assumption does not always hold true \cite{Caruso2014}. Generally, $\mathcal{E}$ might retain partial memory of its interaction with the sequence of qubits passing through it. Consequently, classical correlations can emerge between consecutive uses of the channel. Macchiavello and Palma \cite{Palma2002} proposed a correlated quantum channel model where the Kraus operators $E_{ij}$ in Eq. \eqref{Kraus1} are replaced by $E_{i j}=\sqrt{p_{ij}} \sigma_i \otimes \sigma_j$, incorporating the joint probability
\begin{equation}\label{pij}
p_{i j}=(1-\mu) p_i p_j+\mu p_i \delta_{i j},
\end{equation}
in which $\delta_{i j}$ represents the Kronecker delta function, and the parameter $\mu \in [0, 1]$ quantifies the extent of classical correlations between consecutive operations of the channel $\mathcal{E}$ on the two qubits. It is worth mentioning that this model encompasses the typical scenario of an uncorrelated dephasing channel.

To clarify, we specifically examine the dephasing channel characterized by the probability distribution: $p_0=1-p$, $p_{1, 2}=0$, and $p_3=p$ (see Ref. \cite{Hu2019}). Additionally, for investigating the temporal evolution of quantum coherence, we introduce a colored pure dephasing model governed by a time-varying Hamiltonian $\hat{H}=\hbar\Gamma(t)\sigma_z$ \cite{Daffer2004}. Here, $\Gamma(t)$ represents an independent random variable following the statistics of a random telegraph signal defined by $\Gamma(t)=\beta n(t)$, where $n(t)$ follows a Poisson distribution with $\langle n(t)\rangle=t/2\tau$, and $\beta$ is a coin-flip random variable with possible values $\pm\beta$. When $\beta=1$, the time-dependent parameter $p$ can be derived as $p=[1-f(t)]/2$.

In the non-Markovian regime, namely $\tau>1/4$, the function $f(t)$ is defined by \cite{Hu2019}
\begin{equation}\label{ft1}
f(t)=e^{-t/2\tau}\left[\cos \frac{v t}{2 \tau}+\frac{1}{v} \sin \frac{v t}{2 \tau}\right],
\end{equation}
with $v=\sqrt{|1-16\tau^2|}$. However, in the Markovian regime, i.e. $\tau<1/4$, $f(t)$ takes the following form
\begin{equation}\label{ft2}
f(t)=e^{-t/2\tau}\left[\cosh \frac{v t}{2 \tau}+\frac{1}{v} \sinh \frac{v t}{2 \tau}\right].
\end{equation}

Now, we can derive the thermal time-dependent density operator for two qubits (gravcats) passing through a correlated dephasing channel. Let us consider the input state $\rho_T$ defined in \eqref{eq4}, which is the thermal density operator of gravcats. Then, using Eqs. \eqref{Kraus1} and \eqref{pij}, the following output state would be given by

\begin{equation}\label{rhott}
\rho_T (t)=\left(\begin{array}{cccc}
a^- & 0 & 0 & \eta c \\
0 & b & \eta d & 0\\
0 & \eta d & b & 0\\
\eta c & 0 & 0 & a^+
\end{array}\right),
\end{equation}
where $\eta=f^2(t) + [1-f^2(t)]\mu$.

Note that we have $\eta=1$ for a fully correlated dephasing channel, i.e. $\mu=1$. So, the density operator \eqref{rhott} will be time-independent in this particular case (gravcats).

\begin{figure*}[t]
\centering
{\includegraphics[width=80mm]{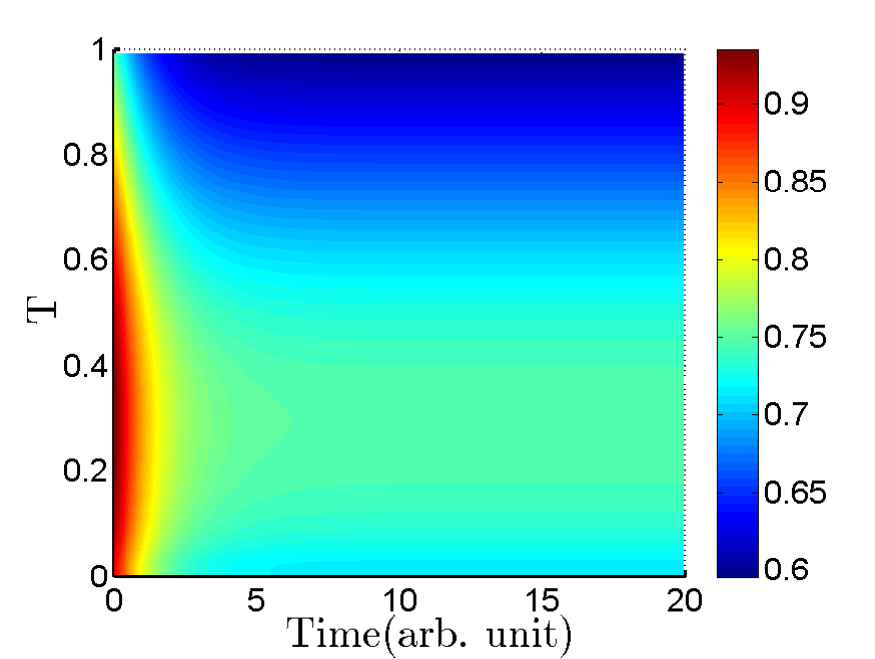}}
\put(-185,162){(a)}\
{\includegraphics[width=80mm]{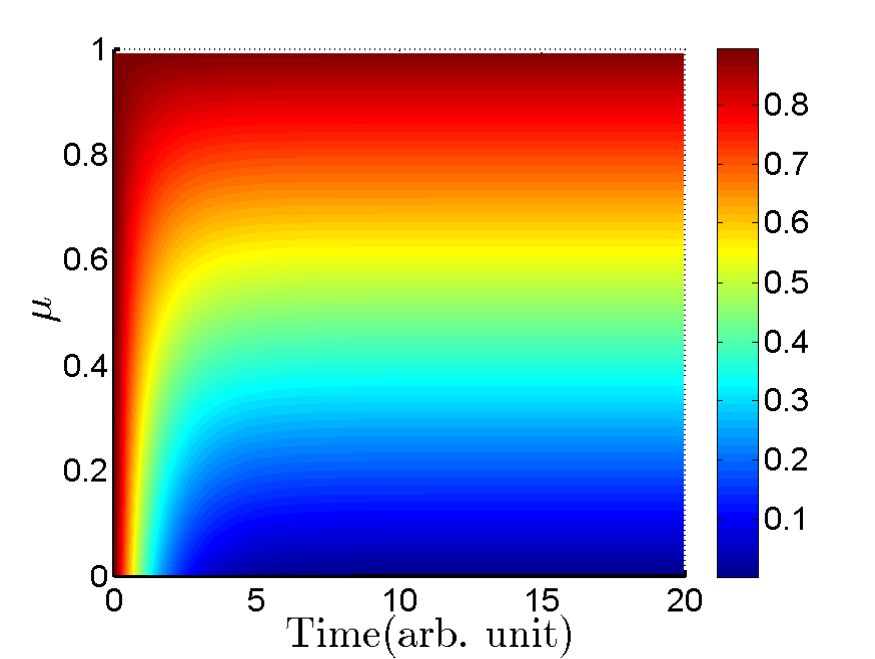}}
\put(-185,162){(b)}\\
{\includegraphics[width=80mm]{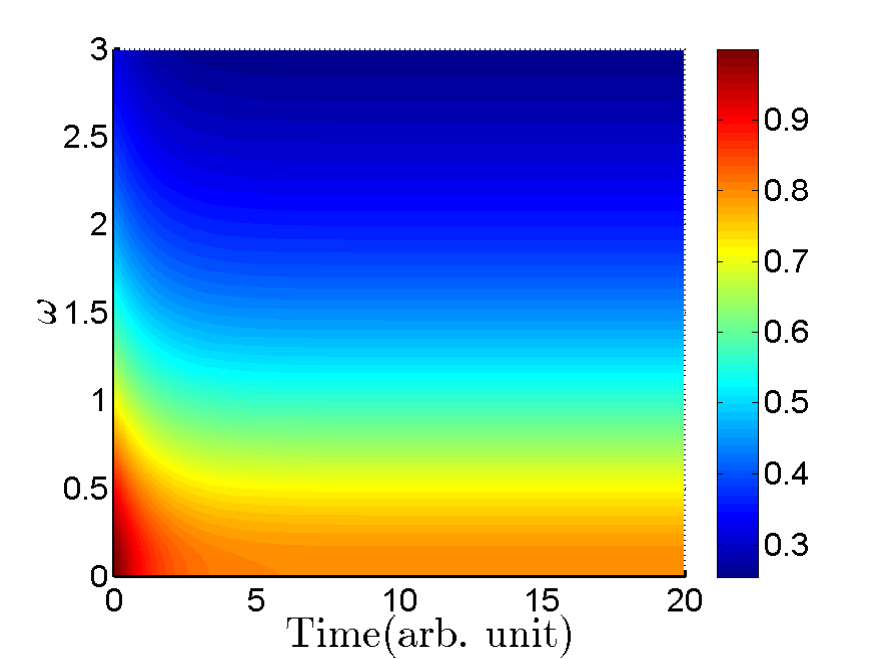}}
\put(-185,162){(c)}\
{\includegraphics[width=80mm]{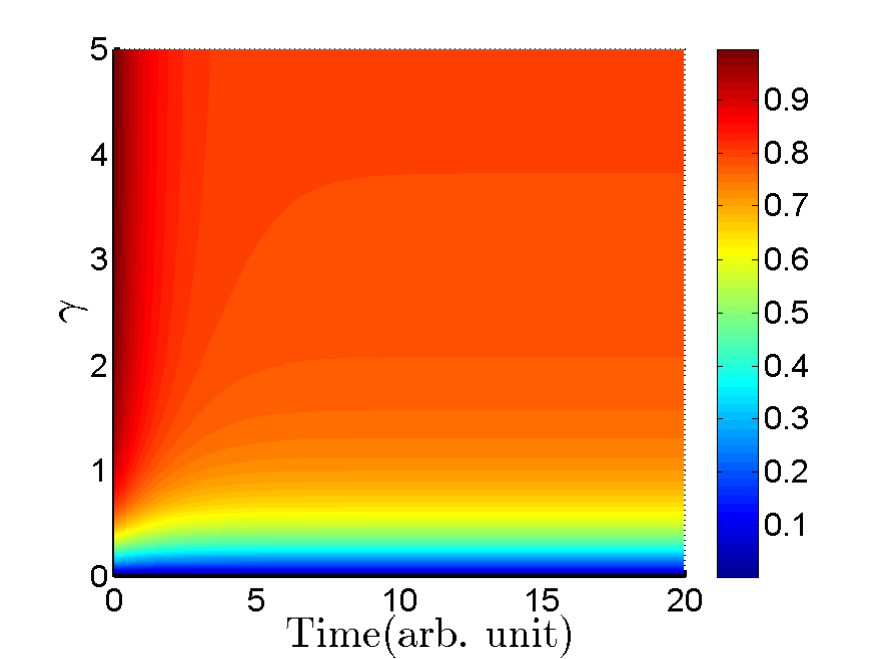}}
\put(-185,162){(d)}
\caption{Time dependence of quantum coherence $C_{l_{1}}$ in the Markovian regime with $\tau=0.1$ for (a) $2\omega=\gamma=1$; $\mu=0.8$,  (b) $2\omega=\gamma=1$; $T=0.01$, (c) $\gamma=1$;  $T=0.01$; $\mu=0.8$, and (d) $\omega=0.5$;  $T=0.01$; $\mu=0.8$.} \label{figure2}
\end{figure*}

\begin{figure*}[t]
\centering
{\includegraphics[width=80mm]{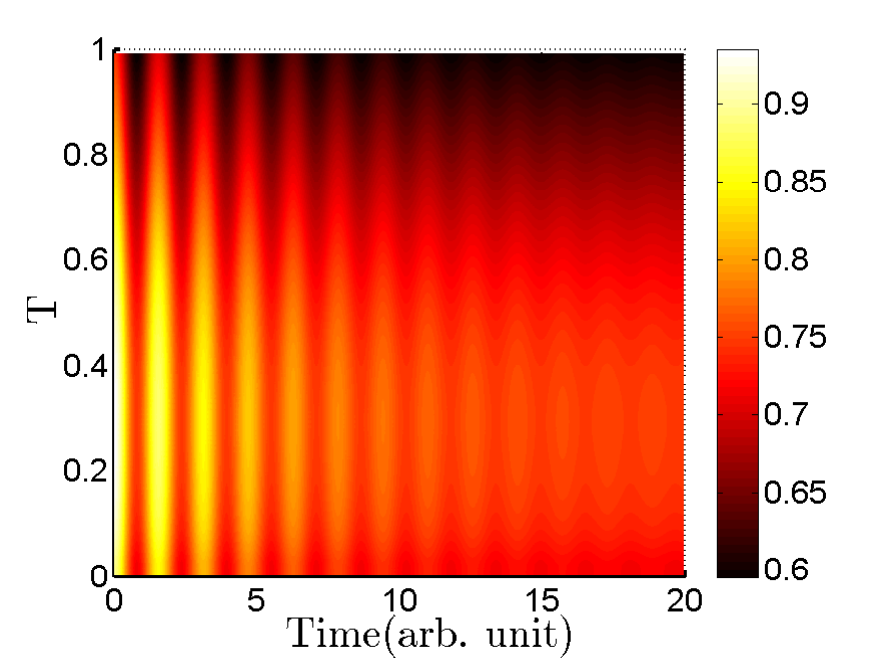}}
\put(-185,162){(a)}\
{\includegraphics[width=80mm]{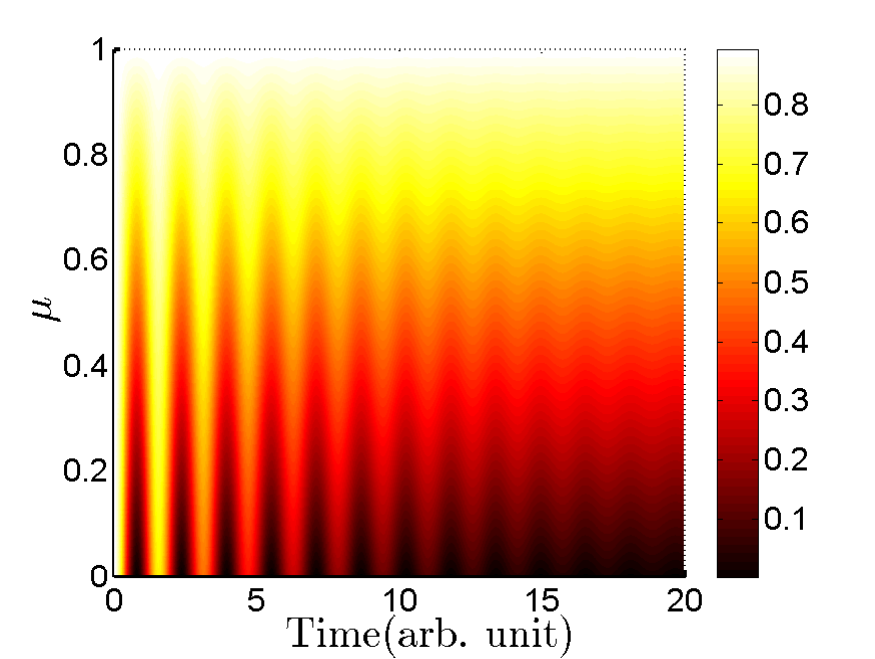}}
\put(-185,162){(b)}\\
{\includegraphics[width=80mm]{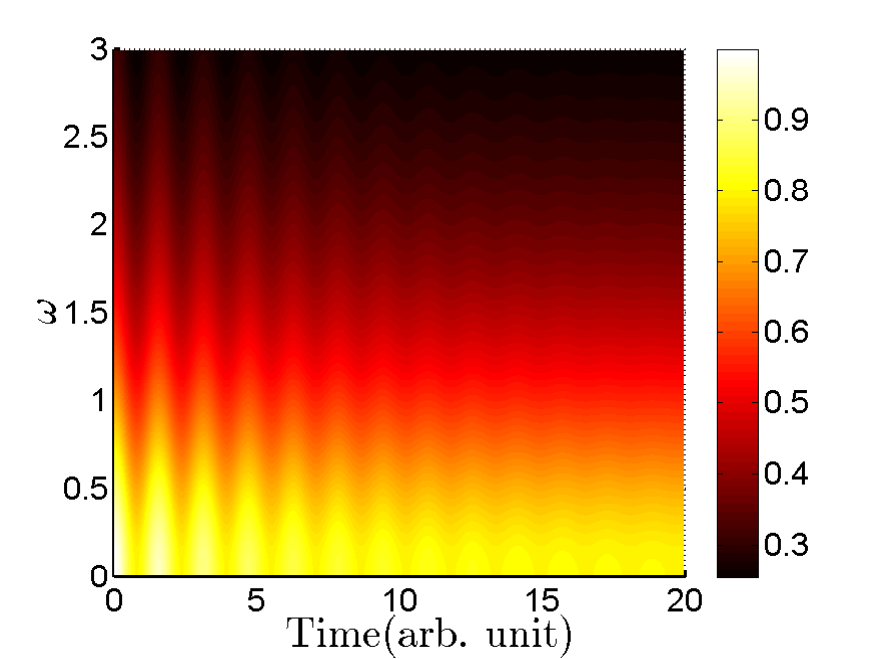}}
\put(-185,162){(c)}\
{\includegraphics[width=80mm]{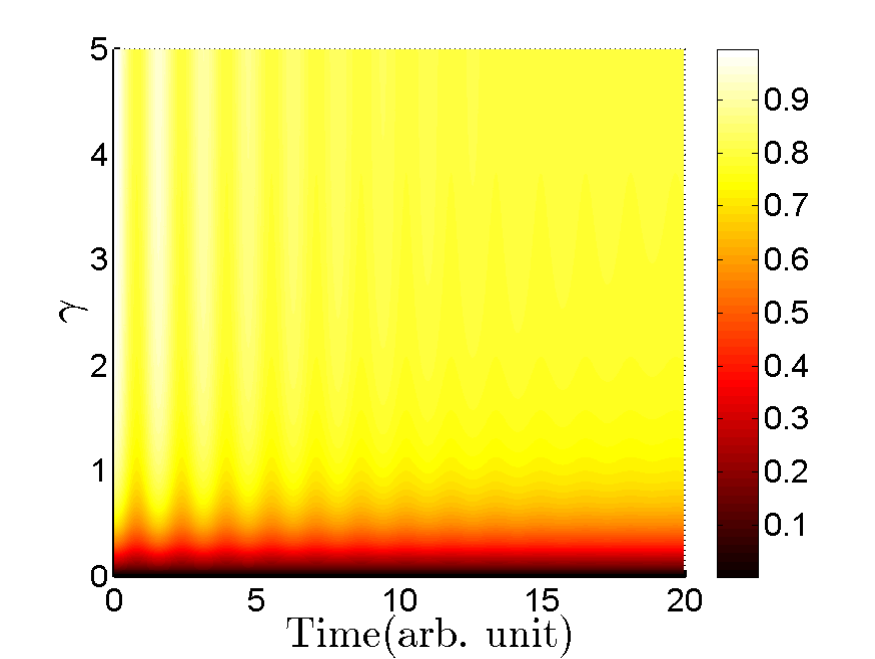}}
\put(-185,162){(d)}
\caption{Time dependence of quantum coherence $C_{l_{1}}$ in the non-Markovian regime with $\tau=5$ for (a) $2\omega=\gamma=1$; $\mu=0.8$,  (b) $2\omega=\gamma=1$; $T=0.01$, (c) $\gamma=1$;  $T=0.01$; $\mu=0.8$, and (d) $\omega=0.5$;  $T=0.01$; $\mu=0.8$.} \label{figure3}
\end{figure*}

\begin{figure*}[t]
\centering
{\includegraphics[width=80mm]{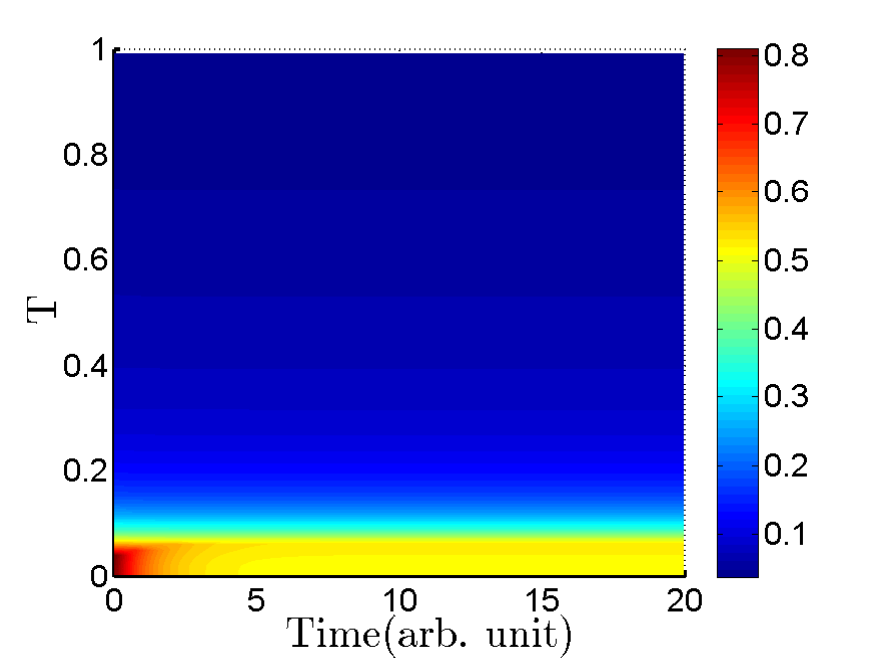}}
\put(-185,162){(a)}\
{\includegraphics[width=80mm]{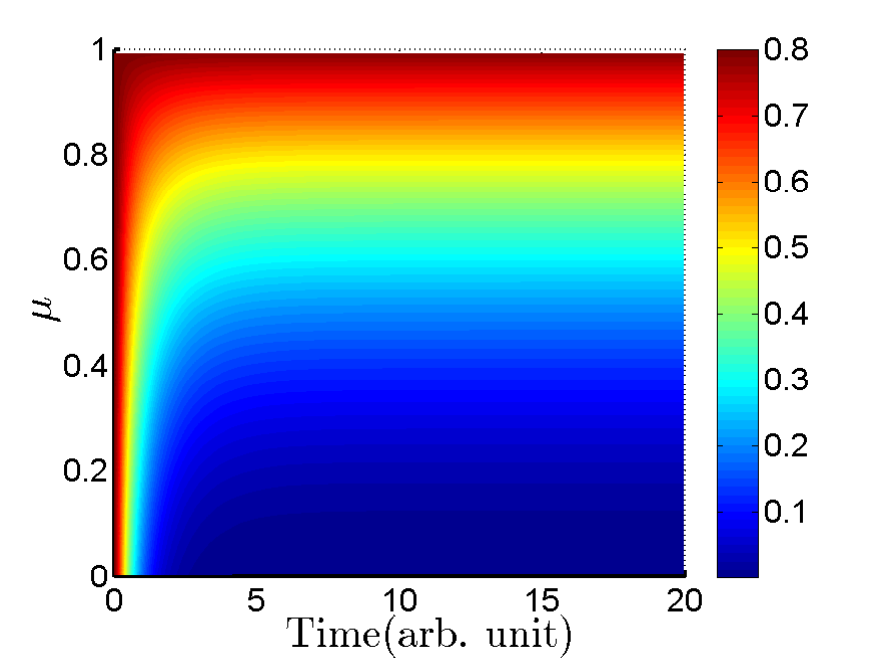}}
\put(-185,162){(b)}\\
{\includegraphics[width=80mm]{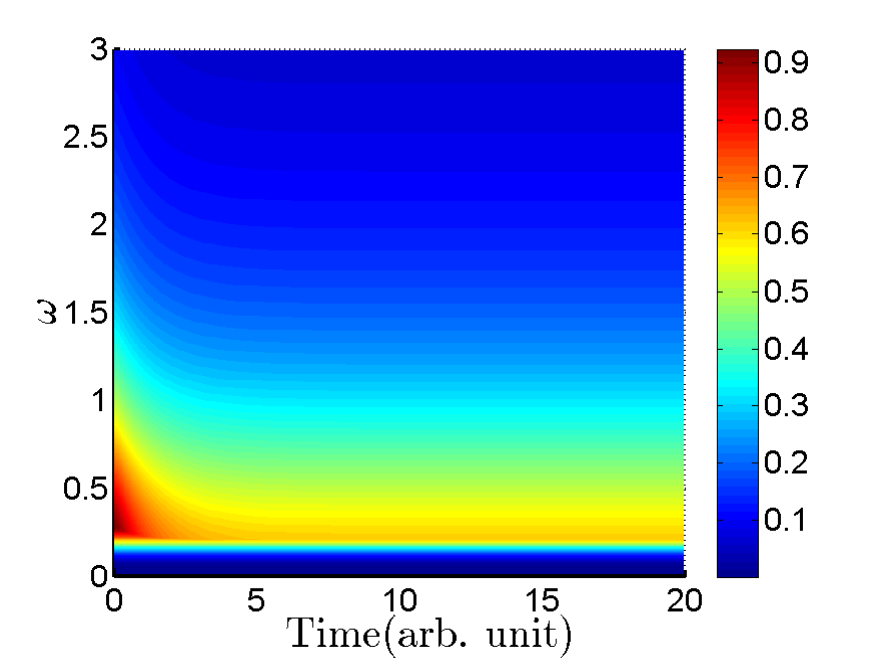}}
\put(-185,162){(c)}\
{\includegraphics[width=80mm]{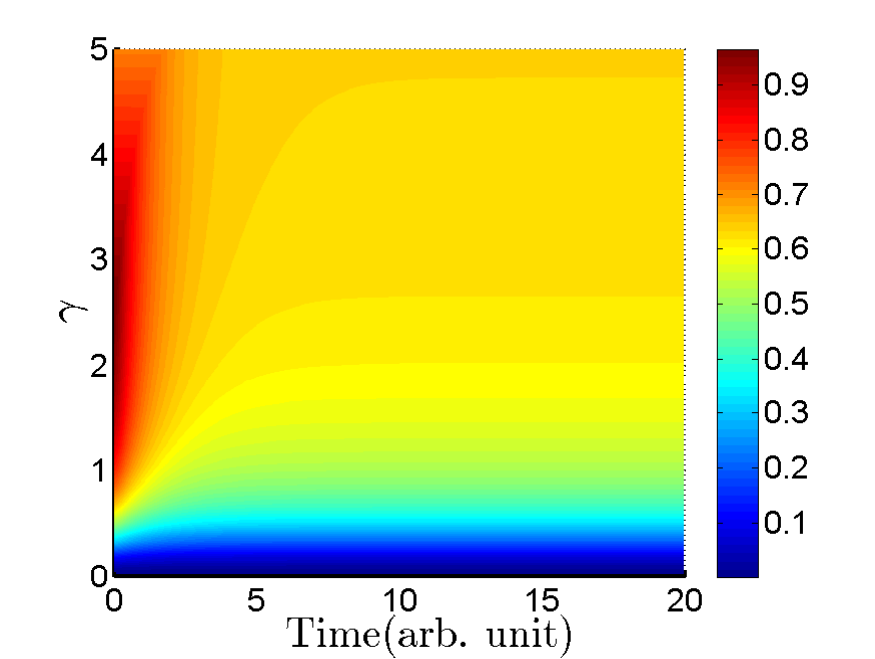}}
\put(-185,162){(d)}
\caption{Time dependence of LQFI $\mathcal{F}$ in the Markovian regime with $\tau=0.1$ for (a) $2\omega=\gamma=1$; $\mu=0.8$,  (b) $2\omega=\gamma=1$; $T=0.01$, (c) $\gamma=1$;  $T=0.01$; $\mu=0.8$, and (d) $\omega=0.5$;  $T=0.01$; $\mu=0.8$.} \label{figure4}
\end{figure*}

\begin{figure*}[t]
\centering
{\includegraphics[width=80mm]{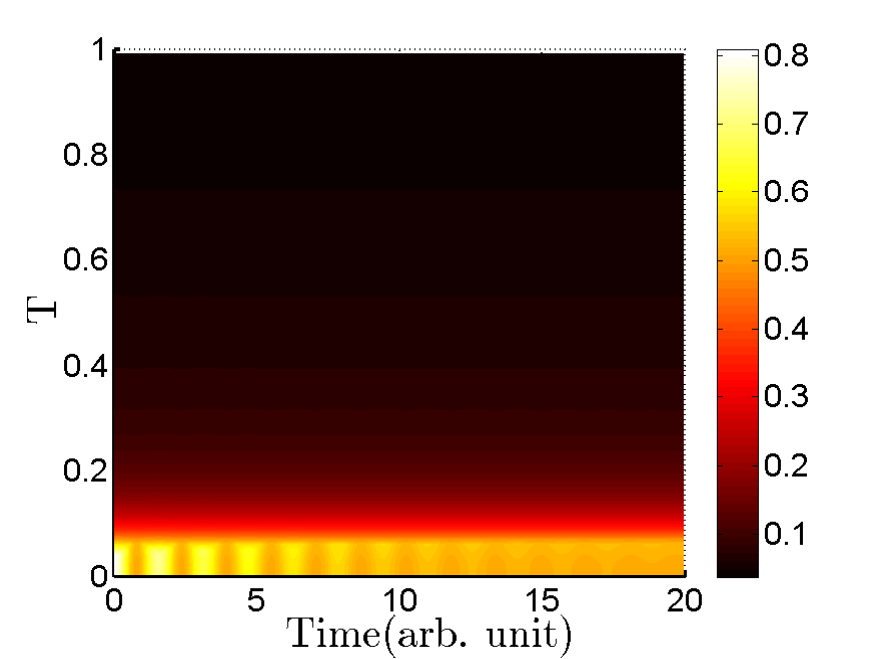}}
\put(-185,162){(a)}\
{\includegraphics[width=80mm]{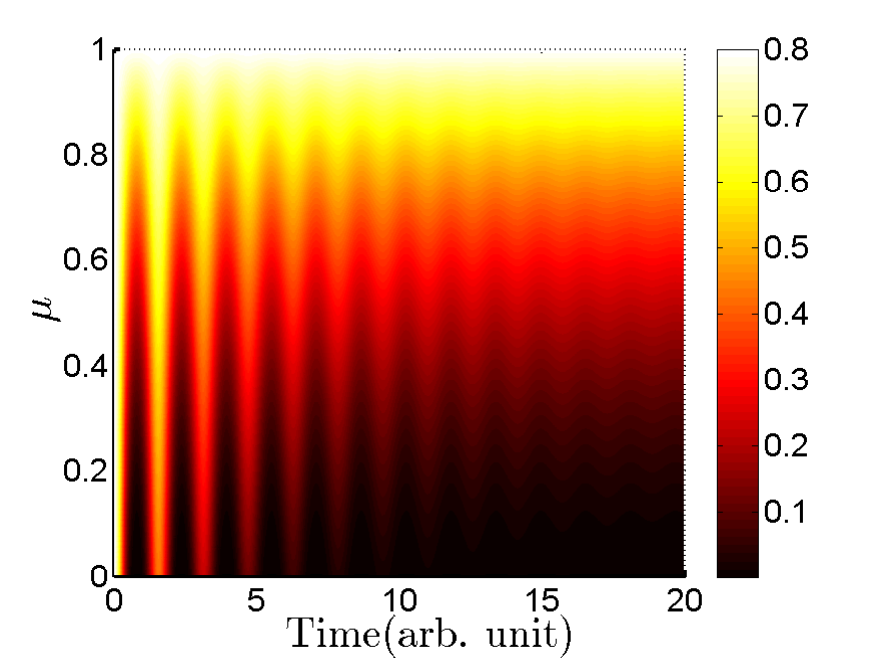}}
\put(-185,162){(b)}\\
{\includegraphics[width=80mm]{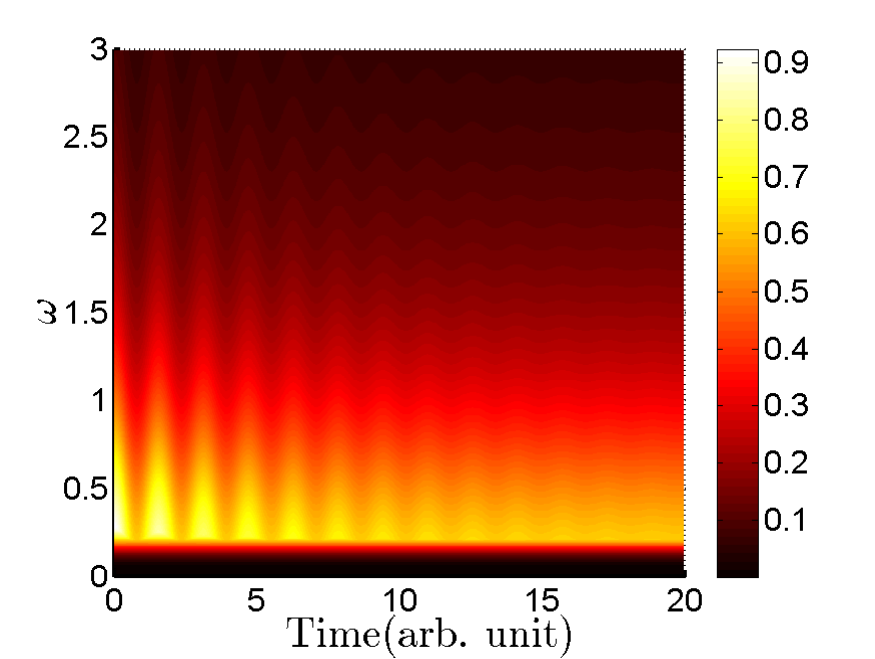}}
\put(-185,162){(c)}\
{\includegraphics[width=80mm]{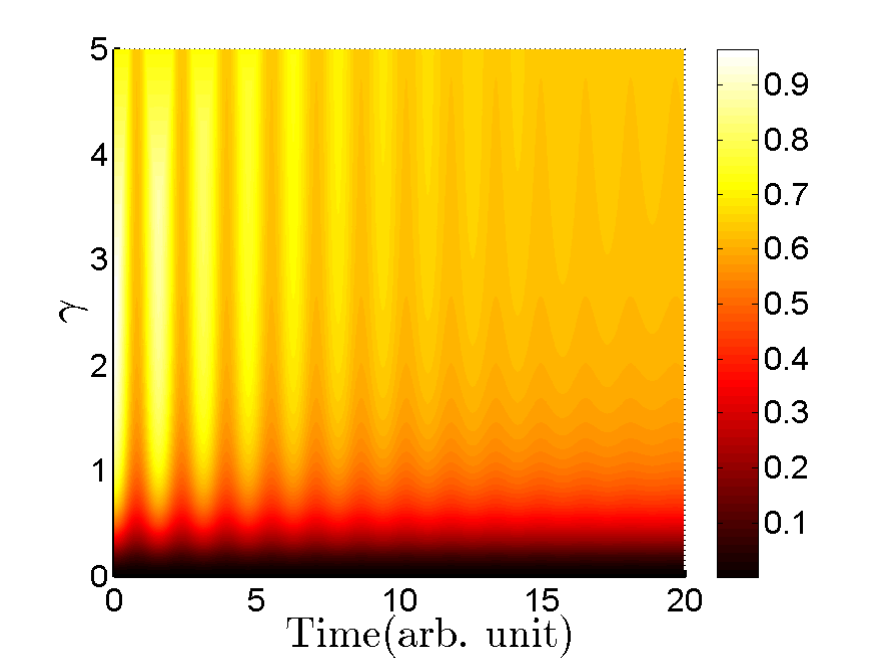}}
\put(-185,162){(d)}
\caption{Time dependence of LQFI $\mathcal{F}$ in the non-Markovian regime with $\tau=5$ for (a) $2\omega=\gamma=1$; $\mu=0.8$,  (b) $2\omega=\gamma=1$; $T=0.01$, (c) $\gamma=1$;  $T=0.01$; $\mu=0.8$, and (d) $\omega=0.5$;  $T=0.01$; $\mu=0.8$.} \label{figure5}
\end{figure*}

\begin{figure*}[t]
\centering
{\includegraphics[width=80mm]{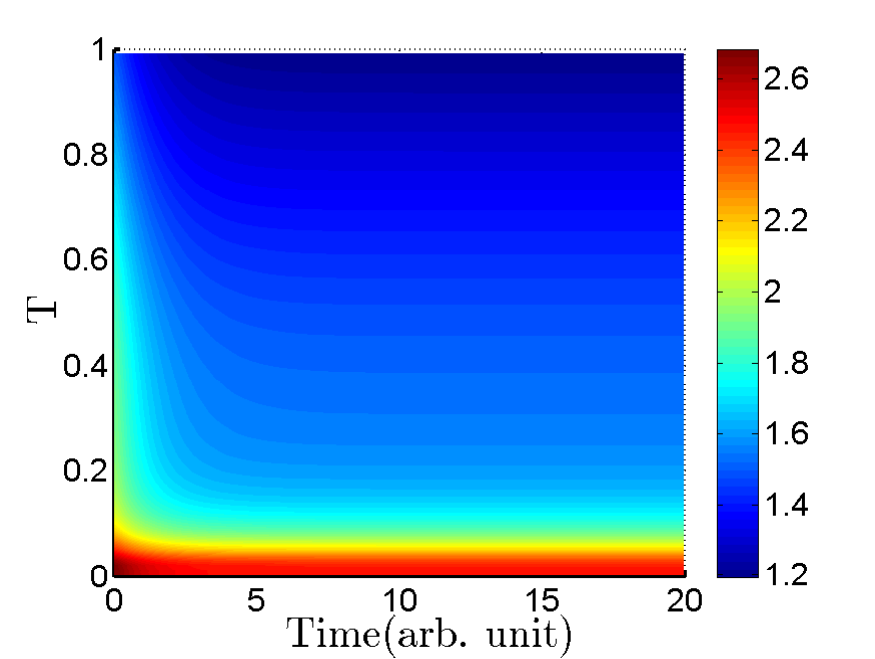}}
\put(-185,162){(a)}\
{\includegraphics[width=80mm]{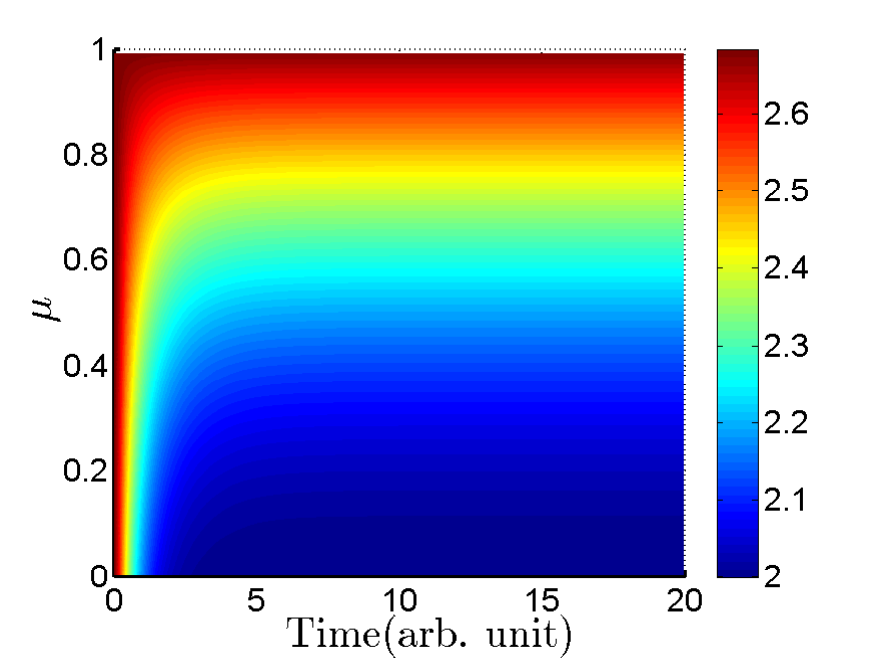}}
\put(-185,162){(b)}\\
{\includegraphics[width=80mm]{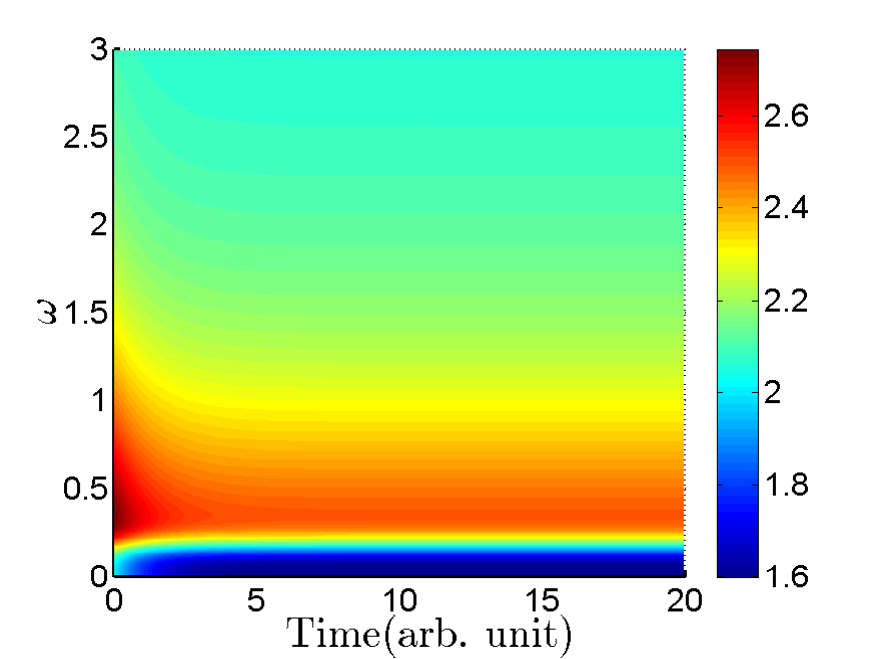}}
\put(-185,162){(c)}\
{\includegraphics[width=80mm]{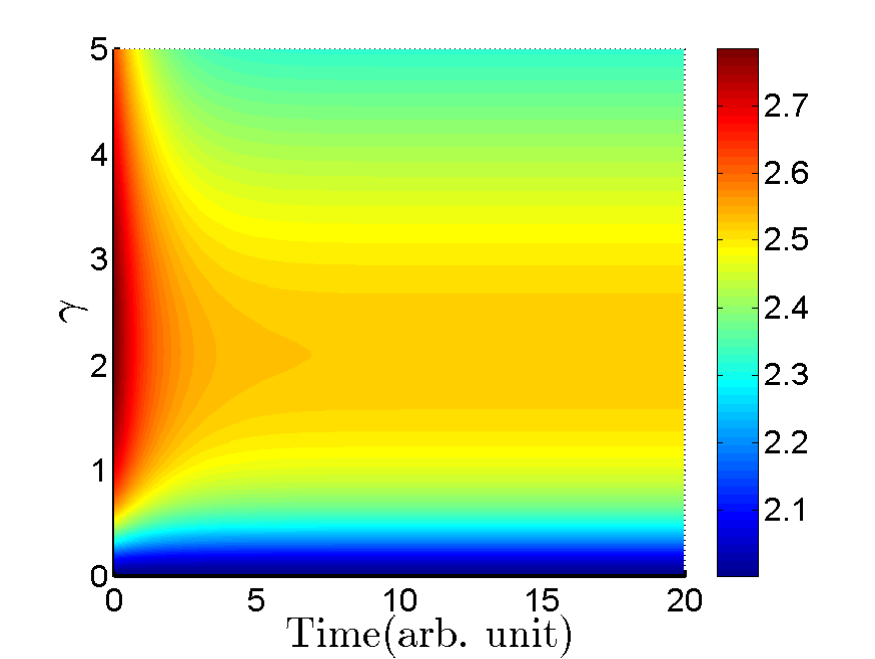}}
\put(-185,162){(d)}
\caption{Time dependence of Bell non-locality $B_{\max}$ in the Markovian regime with $\tau=0.1$ for (a) $2\omega=\gamma=1$; $\mu=0.8$,  (b) $2\omega=\gamma=1$; $T=0.01$, (c) $\gamma=1$;  $T=0.01$; $\mu=0.8$, and (d) $\omega=0.5$;  $T=0.01$; $\mu=0.8$.} \label{figure6}
\end{figure*}

\begin{figure*}[t]
\centering
{\includegraphics[width=80mm]{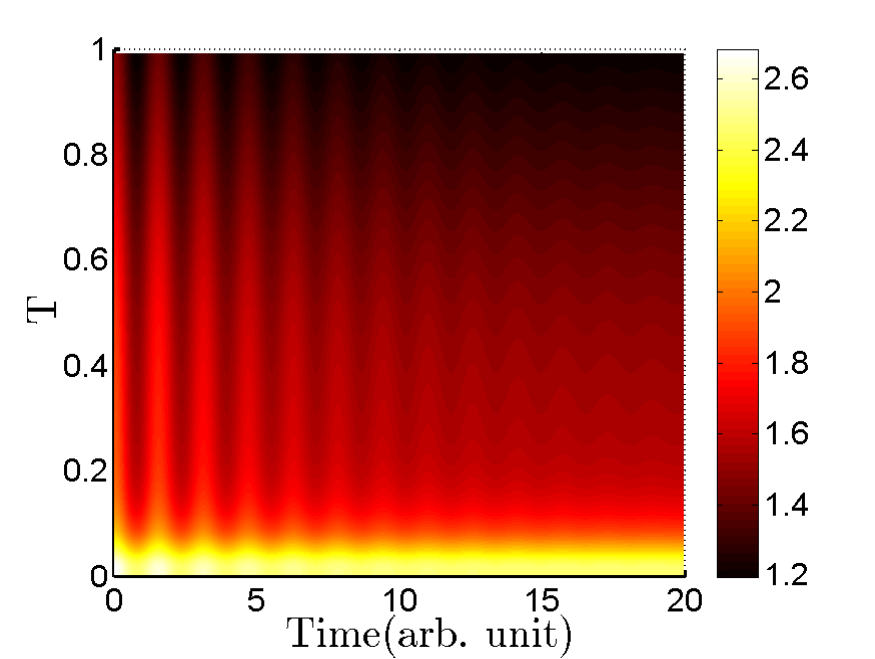}}
\put(-185,162){(a)}\
{\includegraphics[width=80mm]{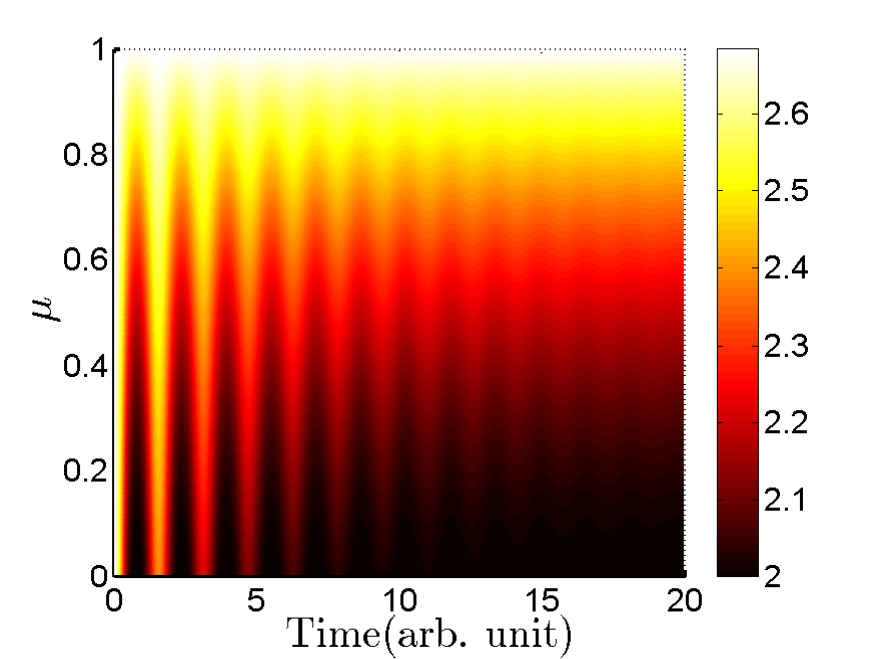}}
\put(-185,162){(b)}\\
{\includegraphics[width=80mm]{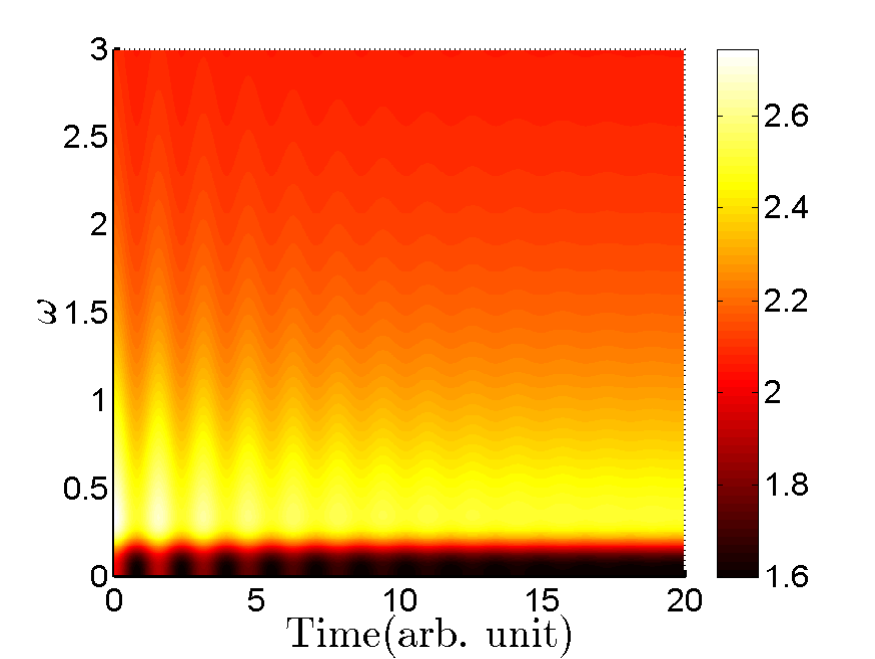}}
\put(-185,162){(c)}\
{\includegraphics[width=80mm]{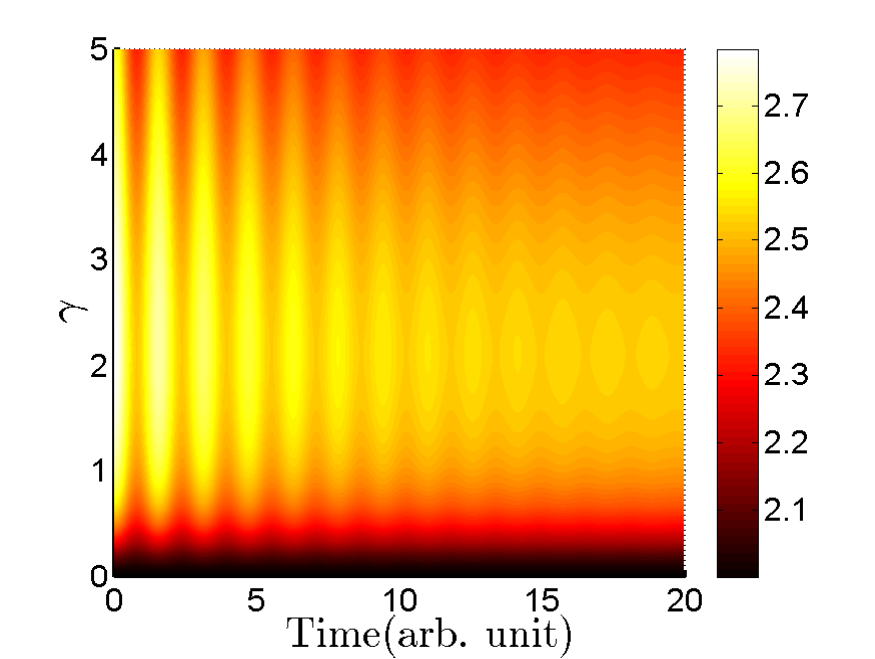}}
\put(-185,162){(d)}
\caption{Time dependence of Bell non-locality $B_{\max}$ in the non-Markovian regime with $\tau=5$ for (a) $2\omega=\gamma=1$; $\mu=0.8$,  (b) $2\omega=\gamma=1$; $T=0.01$, (c) $\gamma=1$;  $T=0.01$; $\mu=0.8$, and (d) $\omega=0.5$;  $T=0.01$; $\mu=0.8$.} \label{figure7}
\end{figure*}

\begin{figure*}[t]
\centering
{\includegraphics[width=50mm]{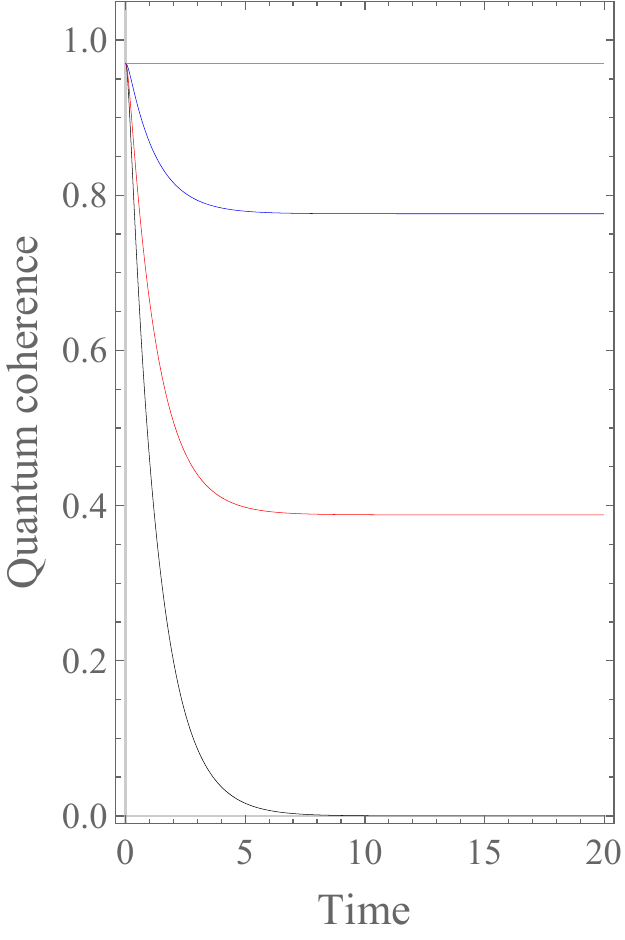}}
\put(-105,218){(a)}\qquad
{\includegraphics[width=50mm]{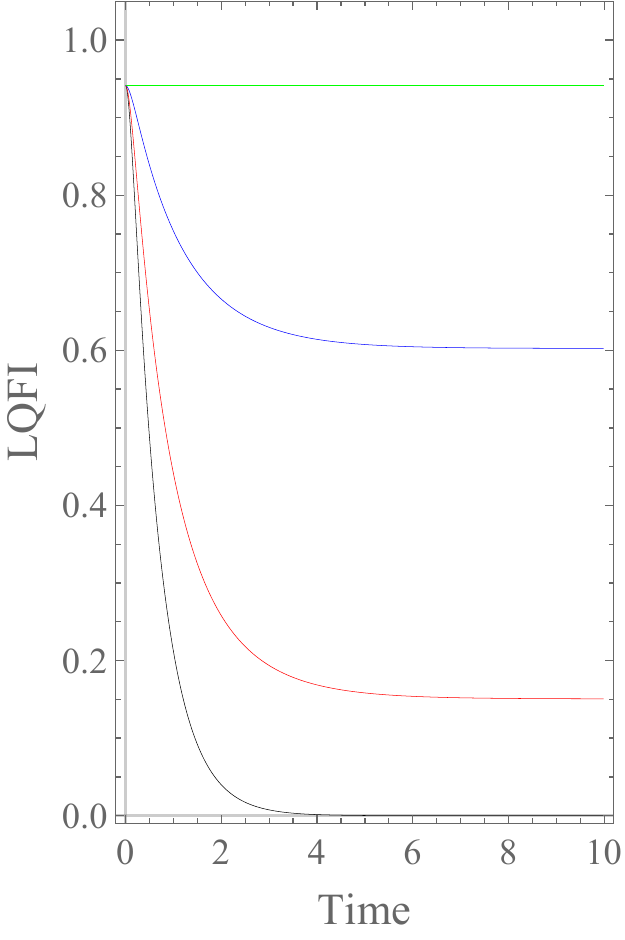}}
\put(-105,218){(b)}\qquad
{\includegraphics[width=51mm]{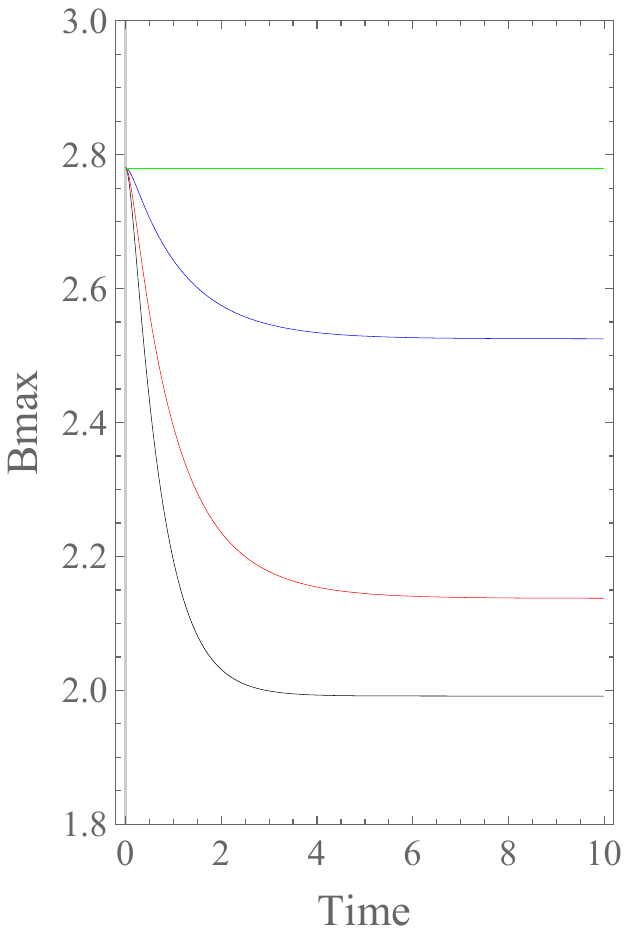}}
\put(-105,218){(c)}
\caption{Time dependence of quantum coherence (a), LQFI (b), and Bell non-locality (c) in the Markovian regime with $\tau=0.1$  for  $4\omega=\gamma=2$ and $T=0.01$. For three plots, the black, red, blue, and green curves
correspond to $\mu = 0, 0.4, 0.8$, and 1, respectively.} \label{figure8}
\end{figure*}

\begin{figure*}[t]
\centering
{\includegraphics[width=50mm]{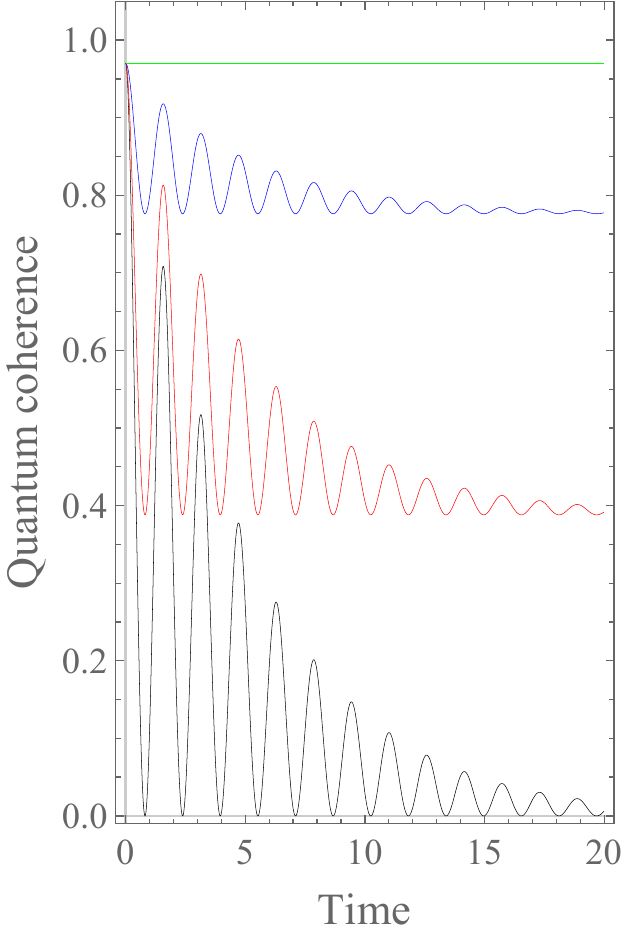}}
\put(-105,218){(a)}\qquad
{\includegraphics[width=50mm]{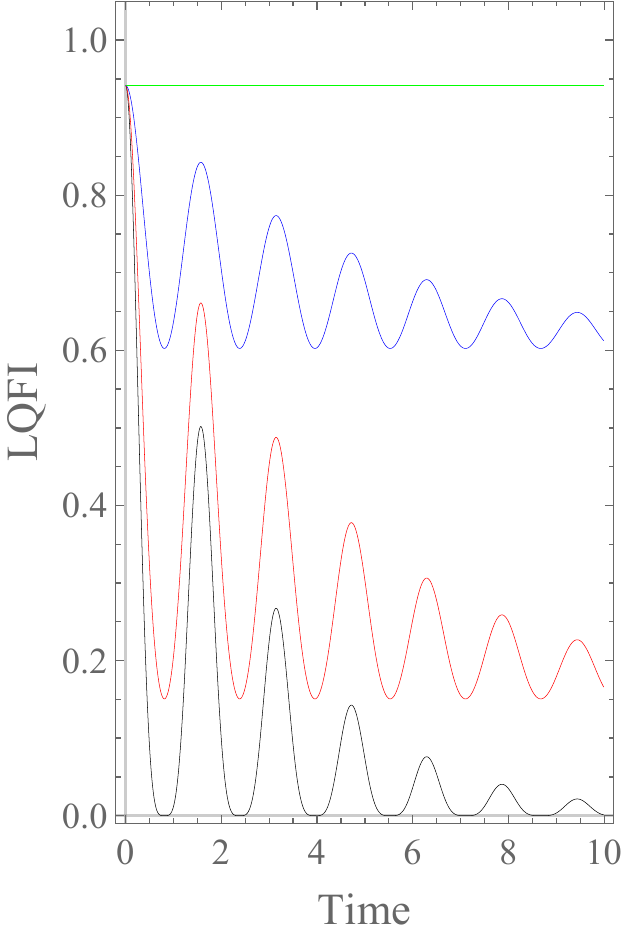}}
\put(-105,218){(b)}\qquad
{\includegraphics[width=51mm]{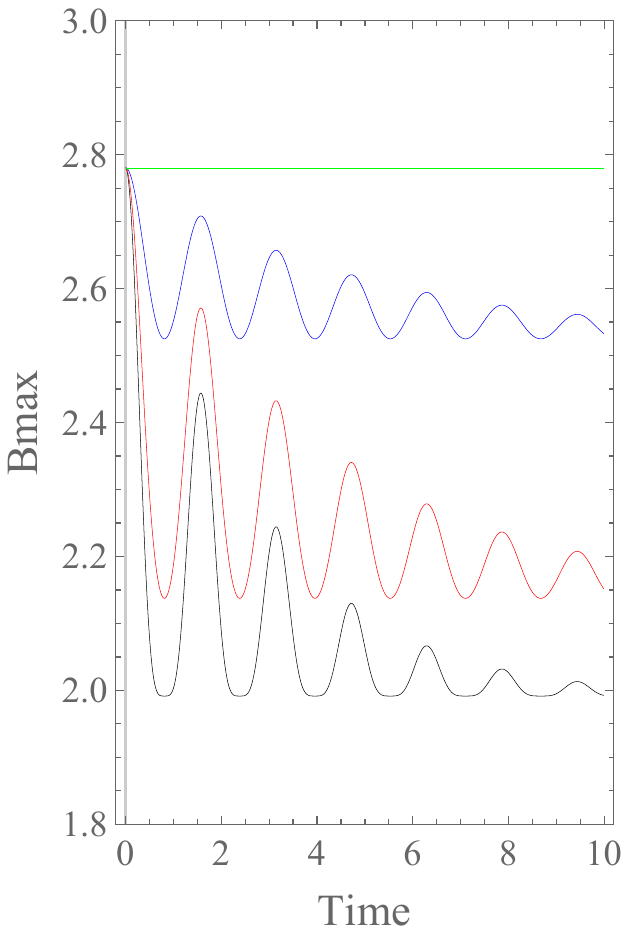}}
\put(-105,218){(c)}
\caption{Time dependence of quantum coherence (a), LQFI (b), and Bell non-locality (c) in the non-Markovian regime with $\tau=5$  for  $4\omega=\gamma=2$ and $T=0.01$. For three plots, the black, red, blue, and green curves
correspond to $\mu = 0, 0.4, 0.8$, and 1, respectively.} \label{figure9}
\end{figure*}

\begin{figure*}[t]
\centering
{\includegraphics[width=50mm]{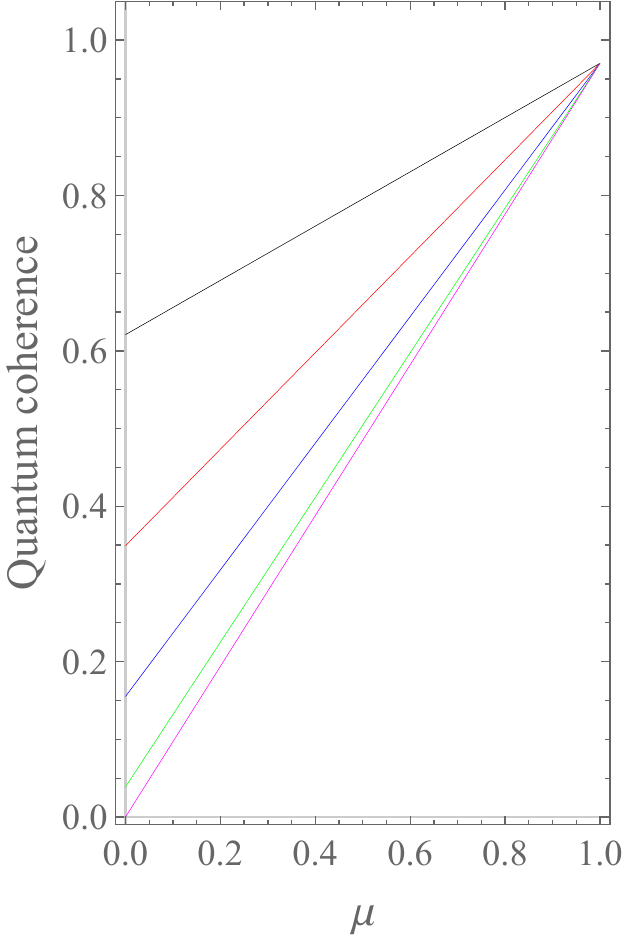}}
\put(-105,218){(a)}\qquad
{\includegraphics[width=50mm]{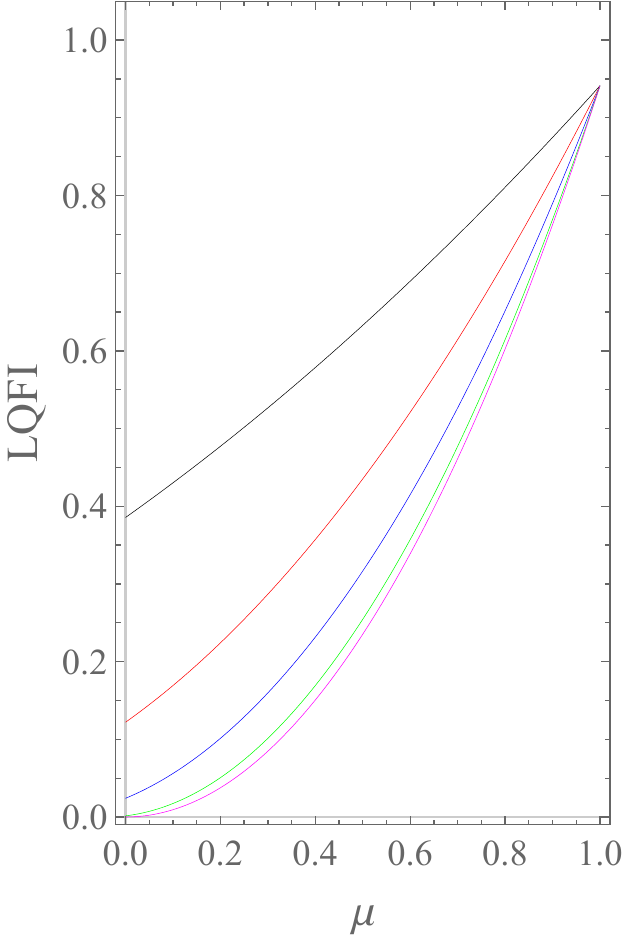}}
\put(-105,218){(b)}\qquad
{\includegraphics[width=51mm]{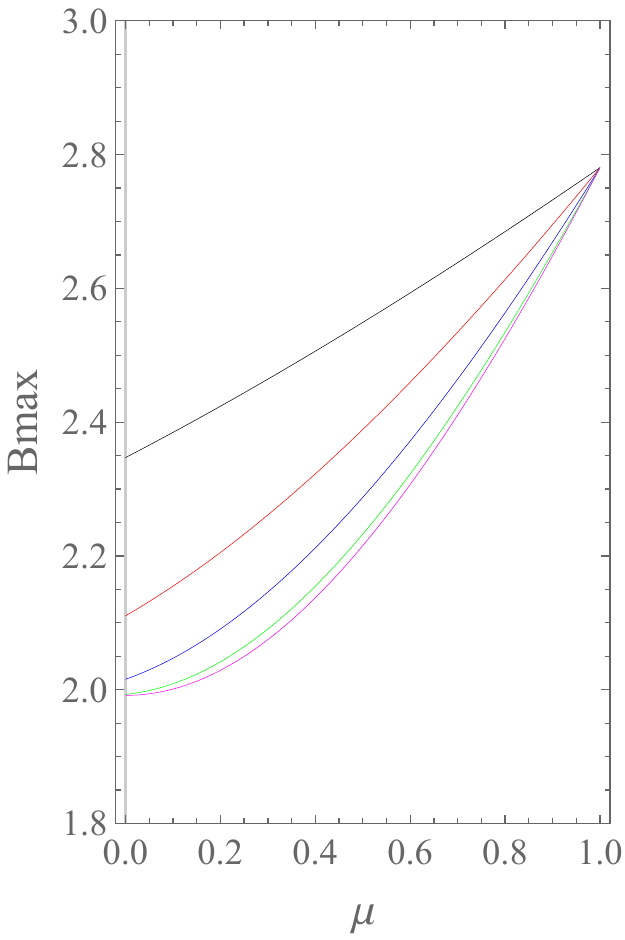}}
\put(-105,218){(c)}
\caption{ The $\mu$ dependence of quantum coherence (a), LQFI (b), and Bell non-locality (c)  with fixed $f=0.8, 0.6, 0.4, 0.2$, and $0$  (from top to bottom).} \label{figure10}
\end{figure*}

\section{Quantum coherence of gravcats}\label{sec:5}

Quantum coherence characterizes the state of quantum systems wherein diverse states coexist simultaneously through superposition \cite{Baumgratz2014}. This property is vital for various quantum technologies like computing and cryptography. In a coherent state, systems can demonstrate interference effects, facilitating accurate management and manipulation of their behavior. Nonetheless, coherence is delicate and susceptible to disturbance from environmental interactions, resulting in decoherence. Consequently, sustaining coherence presents a notable hurdle in the advancement of practical quantum technologies.

Several methods exist for assessing quantum coherence \cite{Streltsov2017}. Among these, the $l_{1}$-norm of coherence is extensively used in quantum physics and is expressed as follows
\begin{equation}\label{Quantumcoherence}
C_{l_{1}}[\rho]=\sum_{i\neq j} |\rho_{i,j}|.
\end{equation}

According to Eq. \eqref{Quantumcoherence}, the $l_{1}$-norm of coherence value can be represented as the sum of absolute values of the off-diagonal elements corresponding to the chosen basis \cite{mlhu2018}.

Using the system's state given in Eq. \eqref{rhott}, we arrive at a formula for the $l_{1}$-norm of coherence as follows
\begin{equation}\label{QC1}
C_{l_{1}}[\rho_T (t)]=2\left(|\eta c|+|\eta d|\right).
\end{equation}

Notice that based on Ref. \cite{Pei2012},  quantum coherence presents the quantum consonance for an X state like our case in \eqref{rhott}.

In Fig. \ref{figure2}, we present the quantum coherence \eqref{QC1} in the Markovian regime versus time and one of the other parameters: $T$, $\mu$, $\omega$ or $\gamma$. In particular, Fig. \ref{figure2} (a) shows $C_{l_{1}}$ in the domain of time and temperature. We observe a decrease in coherence over time, with a faster decay rate at higher temperatures since at lower temperatures, coherence is better preserved due to reduced thermal effects. Generally, from Fig. \ref{figure2} (a), we can conclude that as temperature increases, thermal fluctuations become more pronounced, leading to enhanced decoherence.

Increasing $\mu$ implies stronger classical correlations between successive operations, which can influence the coherence dynamics. Therefore, in Fig. \ref{figure2} (b), we present $C_{l_{1}}$ versus time and $\mu$. One observes slower decay rates of coherence with increasing $\mu$, which proves that classical correlations effectively delay decoherence.

In Fig. \ref{figure2} (c), we observe the system's behavior in response to the energy gap, $\omega$, and time. The lower initial coherence for higher energy gaps is intriguing and suggests that a larger energy gap may hinder the initial establishment of coherence. Despite the differences in initial coherence, the decay rates appear similar for different $\omega$ values, which implies that once coherence is established, the subsequent decay dynamics are predominantly influenced by factors other than the energy gap.

Finally, Fig. \ref{figure2} (d) presents coherence in the Markovian regime versus time and $\gamma$. We obtain higher coherences for greater $\gamma$ values coupled with minimal decay, which indicates the robustness of coherence against gravitational interactions. This also suggests that stronger gravitational interactions might actually stabilize coherence to some extent. It seems that the system reaches a steady state of coherence relatively quickly, and the strength of gravitational interaction does not significantly affect the long-term coherence dynamics.

The results depicted in Fig. \ref{figure2} can be compared with those in Fig. \ref{figure3} where quantum coherence is presented in the same domains but within a non-Markovian regime. Transitioning to a non-Markovian regime introduces additional complexities to coherence dynamics,  including memory effects and non-exponential decay, which manifest as periodic oscillations in quantum coherence.

In each plot from Fig. \ref{figure3}, we observe that quantum coherence does not exhibit monotonic decay over time; instead, it features periodic oscillations, characteristic of non-Markovian effects.

More specifically, in Fig. \ref{figure3} (a), higher temperatures still generally lead to enhanced decoherence due to increased thermal fluctuations. However, the presence of memory effects results in the non-monotonic behavior of quantum coherence over time.

Regarding the strength of classical correlations, Fig. \ref{figure3} (b) demonstrates that the amplitude of oscillations of quantum coherence decreases as the classical correlation parameter, $\mu$, increases. Additionally, the decay rate is significantly slower for higher $\mu$ values, supporting the hypothesis that stronger classical correlations lead to better preservation of quantum coherence.

In the non-Markovian regime, the energy gap between states still influences coherence dynamics, which is presented in Fig. \ref{figure3} (c). While a larger energy gap leads to a lower initial value of coherence, memory effects introduce non-trivial dynamics. For lower energy gaps, the presence of memory effects leads to oscillatory behavior and long-term coherence preservation due to slower decay rates.

Figure \ref{figure3} (d) illustrates the influence of gravitational interactions on coherence dynamics in the non-Markovian regime. Strong gravitational interactions induce memory effects, leading to slower decay rates and oscillatory behavior with smaller amplitudes. Quantum coherence is well-preserved over time for sufficiently large values of $\gamma$.

\section{LQFI of gravcats}
Quantum Fisher information (QFI) originates from quantum metrology, focusing on measurement precision in the quantum domain \cite{qfi1,qfi2,qfi3,qfi5}. In quantum mechanics, Fisher information quantifies the information regarding a parameter contained in measurement outcomes, indicating how well a quantum state can be distinguished from nearby states concerning the measured parameter \cite{qfi7}.

Another utilization of QFI involves employing it as a measure to quantify quantum correlations. Hence, Girolami et al. \cite{qfi7} showed that local quantum Fisher information (LQFI), a computable measure of discord-type quantum correlations, can be expressed as follows
\begin{equation}
\mathcal{F}(\rho):=\min _{H_A} F\left(\rho, H_A\right).
\end{equation}

This discord-type measure (denoted by $\mathcal{F}$) builds upon QFI ($F$) and represents the optimal LQFI when $H_A$ (a measurement operator) is applied to subsystem $A$ belonging to bipartite system $AB$.

In particular, if subsystem $A$ represents a qubit, the following formula can be employed to express the LQFI
\begin{equation}\label{lqfi2}
\mathcal{F}=1-\lambda_{\max}^{\mathcal{M}},
\end{equation}
here, $\lambda_{\max}^{\mathcal{M}}$ represents the maximum eigenvalue of the symmetric 3-by-3 matrix $\mathcal{M}$, whose elements are
\begin{equation}
\mathcal{M}_{\nu \mu}=\sum_{i,j; q_i + q_j \neq 0} \frac{2 q_i q_j}{q_i + q_j} \langle \psi_i |\sigma_\nu \otimes \mathbf{I}|\psi_j\rangle \langle \psi_j |\sigma_\mu \otimes \mathbf{I}|\psi_i\rangle,
\end{equation}
in which $q_i$ and $|\psi_i\rangle$ denote the eigenvalues and eigenstates of $\rho$, given by $\rho=\sum_i q_i |\psi_i\rangle \langle\psi_i|$, where $q_i$ are non-negative and $\sum_i q_i=1$. Besides,  $\sigma_\nu$ and $\sigma_\mu$ show the set of Pauli matrices for $\nu,~\mu = x, y, z$.

Using the explicit formula presented in \cite{YurischevPLA2023}, LQFI for the output state \eqref{rhott} would be given by
\begin{equation}\label{LQFI1}
\mathcal{F}[\rho_T (t)]=\min\{\mathcal{F}_0,~\mathcal{F}_1\},
\end{equation}
where
\begin{equation}
\mathcal{F}_0=1-\mathcal{M}_{zz}
\end{equation}
and
\begin{equation}
\mathcal{F}_1=1-\mathcal{M}_{xx},
\end{equation}
with
\begin{equation}
\mathcal{M}_{zz}=1-4\left(\frac{|\eta c|^2}{a^{-} + a^{+}}+\frac{|\eta d|^2}{2b}\right)
\end{equation}
and
\begin{equation}
\mathcal{M}_{xx}=\frac{m_1 m_2}{m_3},
\end{equation}
where
$$
\begin{aligned}
& m_1=64\left(a^{+}b+ a^{-}b+q_1 q_2+q_3 q_4+2\left|\eta^{2} cd\right|\right), \\
& m_2=\left(a^{+}+a^{-}\right) q_3 q_4+2b q_1 q_2
\end{aligned}
$$
and
$$
\begin{aligned}
m_3= & {\left[1-\left(q_1-q_2\right)^2-\left(q_3-q_4\right)^2\right]^2 } \\
& -4\left(q_1-q_2\right)^2\left(q_3-q_4\right)^2,
\end{aligned}
$$
in which $q_i$'s $(i=1,2,3,4)$ are the eigenvalues of our state  \eqref{rhott}.

Figures~\ref{figure4} and \ref{figure5} present LQFI in the Markovian and non-Markovian regimes, respectively. To facilitate comparison, we have investigated LQFI within the same domain as in the case of quantum coherence.

In Fig.~\ref{figure4} (a), LQFI demonstrates a decrease over time in the Markovian regime, with faster decay rates at higher temperatures due to increased thermal effects. The initial values of LQFI vary depending on the temperature, reflecting the degree of coherence in the system at the beginning of the evolution. Comparing these findings with the non-Markovian regime depicted in Fig. \ref{figure5} (a), we see that higher temperatures still lead to increased thermal effects, resulting in enhanced decoherence and suppressed LQFI. However, at lower temperatures in the non-Markovian regime, LQFI displays oscillatory behavior over time, reflecting the influence of memory effects on the coherence evolution.

Figure \ref{figure4} (b) and Fig.~\ref{figure5} (b) demonstrate how classical correlations between consecutive operations of the dephasing channel influence the dynamics of LQFI. Stronger classical correlations, represented by higher values of $\mu$, delay the decay of LQFI, leading to better preservation of quantum advantages. While in Fig.~\ref{figure5} (b) this tendency is still valid, in the non-Markovian regime memory effects play a significant role, leading to a non-exponential decay and complex LQFI dynamics. If we consider classical correlations such that $\mu \rightarrow 1$, we obtain LQFI that is robust against decoherence over time in both regimes.

In the Markovian regime, the energy gap between the ground state and the first excited state significantly influences LQFI dynamics, which is presented in Fig.~\ref{figure4} (c). A larger energy gap generally leads to initially lower values of LQFI, reflecting reduced coherence at the outset. While this tendency also occurs in the non-Markovian regime, it is modulated by memory effects as presented in Fig. \ref{figure5} (c).

The strength of gravitational interaction between masses $m$ affects LQFI dynamics as shown in Fig.~\ref{figure4} (d) for the Markovian regime. In general, higher gravitational interactions lead to higher initial values of LQFI. Moreover, the decay rate of LQFI appears most significant in the initial period, when the value of LQFI declines sharply. However, after this initial decrease, LQFI stabilizes and the decay does not continue. In the non-Markovian regime, which is presented in Fig.~\ref{figure5} (d), the initial value of LQFI is still affected by $\gamma$, but the dynamics of LQFI are non-monotonic due to memory effects.

\section{Bell non-locality in gravcats}\label{sec:6}
Bell non-locality refers to a quantum effect where measurements on entangled particles reveal correlations that contradict classical physics predictions \cite{Bell1964,Brunner2014}. This phenomenon suggests that the behavior of these particles transcends locality, indicating the inadequacy of explanations based on local hidden variables \cite{CHSH1969}

In the realm of $2\times2$ systems, discerning the non-locality of a quantum state $\rho$ involves testing its violation of the Bell-CHSH inequality, originally formulated by Clauser, Horne, Shimony, and Holt \cite{CHSH1969}. This inequality would be expressed as follows
\begin{equation}\label{Bell-CHSH}
|\langle B_{\textmd{CHSH}}\rangle_{\rho}|\leq 2,
\end{equation}
with $\langle B_{\textmd{CHSH}}\rangle_{\rho}=\textmd{tr}[\rho B_{\textmd{CHSH}}]$, where $B_{\textmd{CHSH}}$ represents the Bell operator linked to the quantum CHSH inequality, the maximal value of inequality \eqref{Bell-CHSH}, denoted as $B_{\max}(\rho)=\max |\langle B_{\textmd{CHSH}}\rangle_{\rho}|$, is associated with the quantity $M(\rho)=\max_{i<j}(\omega_i + \omega_j)$. Here, $B_{\max}(\rho)=2\sqrt{M(\rho)}$, with $\omega_i$ ($i=1,2,3$) indicating the eigenvalues of a $3\times3$ matrix $X^\dagger X$. In this context, $X$ represents a positive matrix with elements $x_{\nu \mu}=\textmd{tr}(\rho \sigma_{\nu} \otimes \sigma_{\mu})$ \cite{Horodecki1995}. The violation of inequality \eqref{Bell-CHSH} is distinctly observable only when $M(\rho)$ surpasses 1. Additionally, $M(\rho)$ serves as a practical measure to assess the extent of Bell non-locality violation within a bipartite state \cite{mlhuqinp2013,Zidan2023}.

For the output state expressed in Eq. \eqref{rhott}, one can obtain
\begin{equation}\label{bell0}
B_{\max}[\rho_T (t)]=2\sqrt{M[\rho_T (t)]},
\end{equation}
where
\begin{equation}
M[\rho_T (t)]=\max\{M_1,~M_2\},
\end{equation}
with
$$M_1=8(|\eta c|^2 +|\eta d|^2)$$
and
$$M_2=4(|\eta c| +|\eta d|)^2 + (a^{+} + a^{-}- 2b)^2.$$

Figures \ref{figure6} and  \ref{figure7} illustrate the behavior of the Bell parameter \eqref{bell0} in the Markovian and non-Markovian regimes, respectively. The Bell parameter serves as a quantifier of non-classical correlations, exceeding $2$ when such correlations are present.

In Fig. \ref{figure6} (a), we observe a decrease in the Bell parameter over time in the Markovian regime, with faster decay rates at higher temperatures due to heightened thermal effects. Higher initial values of the Bell parameter at lower temperatures reflect the preservation of non-classical correlations at the onset of evolution. Contrasting this with the non-Markovian regime depicted in Fig. \ref{figure7} (a), we still observe enhanced decoherence at higher temperatures, suppressing the Bell non-locality $B_{\max}$. However, at lower temperatures in the non-Markovian regime, the Bell parameter displays oscillatory behavior over time, influenced by memory effects.

Figures \ref{figure6} (b) and  \ref{figure7} (b) demonstrate how classical correlations, represented by $\mu$, influence the dynamics of the Bell parameter. Stronger correlations delay the decay of $B_{\max}$, preserving non-classical correlations. This trend persists in the non-Markovian regime, although memory effects introduce non-exponential decay and complex dynamics. When $\mu \rightarrow 1$, the Bell parameter remains close to its maximum value, indicating robust non-classical correlations over time.

The impact of the energy gap ($\omega$) on the Bell parameter dynamics is shown in Fig. \ref{figure6} (c) for the Markovian regime. First, a larger energy gap leads to higher initial values of the Bell parameter. Then, if $\omega$ exceeds a certain threshold, the initial Bell parameter is decreased,  reflecting reduced non-classical correlations. Overall, from Fig. \ref{figure6} (c), we conclude that there exists a specific interval of $\omega$ that ensures a high initial value of $B_{\max}$ and sufficient preservation of non-classical correlations over time. This trend is also observed in the non-Markovian regime [see Fig. \ref{figure7} (c)], albeit modulated by memory effects.

The strength of gravitational interaction ($\gamma$) affects the Bell non-locality dynamics in the Markovian regime, as seen in Fig. \ref{figure6} (d). Higher interactions result in higher initial values of the Bell parameter. For an initial period of time, the Bell parameter declines sharply but stabilizes thereafter. We see a similar tendency as in Fig. \ref{figure6} (c), which means that within an interval of moderate values of $\gamma$, $B_{\max}$ appears well-preserved over time. In the non-Markovian regime [Fig.\ref{figure7} (d)], the dynamics of the Bell non-locality are non-monotonic due to memory effects, with the initial Bell parameter influenced by $\gamma$ in a similar way as in Markovian regime.

The behavior of the Bell parameter provides valuable insights into the preservation of non-classical correlations in gravcat states subjected to correlated dephasing channels. Its similarities with LQFI underscore the interplay between classical correlations, energy gap, gravitational interaction, and memory effects in shaping the quantum characteristics of the system.

\section{Comparative analysis of quantum resources}\label{sec6}
The presented results reveal that quantum coherence, LQFI, and Bell non-locality can be substantially boosted by increasing classical correlations in gravcat states passing through correlated dephasing channels. In this section, we focus on the details of this matter.

We present an insightful examination of the impact of classical correlations (quantified by $\mu$) on the time evolution of quantum resources in the Markovian regime [look at Fig. \ref{figure8}]. The subfigures (a), (b), and (c) depict quantum coherence, LQFI, and the Bell parameter, respectively, each for four different values of $\mu$.

For $\mu = 1$, as depicted in Fig. \ref{figure8}, all quantities remain constant over time because the density operator \eqref{rhott} is time-independent in this special case. This finding underscores the critical role of classical correlations in preserving quantum resources. Conversely, for $\mu < 1$, exponential decay is observed in all quantities. Lower values of $\mu$ correspond to more pronounced decay, highlighting the detrimental effect of reduced classical correlations on the preservation of quantum resources. In the extreme case of $\mu = 0$, quantum coherence and LQFI asymptotically approach zero, while the Bell parameter converges to $2$, indicating the loss of non-classical correlations.

In Fig. \ref{figure9}, we extend our analysis to the non-Markovian regime. Similar to the Markovian case, for $\mu = 1$, all plots remain constant, demonstrating robustness over time. However, for lower values of $\mu$, oscillatory behavior is observed in all quantities due to memory effects. The amplitude of oscillations decays over time. Interestingly, in the absence of classical correlations ($\mu = 0$), all plots eventually converge to equilibrium values, the same as observed in the Markovian regime.

These findings highlight the crucial role of classical correlations in the preservation of quantum resources. In both Markovian and non-Markovian regimes, higher values of $\mu$ ensure greater stability and longevity of quantum resources over time. Conversely, decreasing classical correlations leads to accelerated decay and eventual loss of non-classical correlations. Furthermore, the emergence of oscillatory behavior in the non-Markovian regime highlights a more complex interplay between classical correlations and memory effects in shaping the time evolution of quantum resources.

For a more detailed examination of the relationship between the classical correlation parameter ($\mu$) and various quantum resources -- quantum coherence, LQFI, and Bell non-locality parameter, we draw Fig. \ref{figure10}. This figure consists of three plots, each representing a different quantum resource. In each plot, there are five curves, corresponding to different time instants [by fixing $f$ according to Eqs. \eqref{ft1} and \eqref{ft2}], with the bottom curve representing the infinite-time limit (i.e., $f = 0$).

The overall trend across all plots indicates that quantum resources generally increase with increasing values of $\mu$. This confirms that higher classical correlations contribute to the preservation of quantum resources. Importantly, in each plot, curves corresponding to earlier times lie above those for later times, demonstrating that as time progresses, quantum resources generally decrease due to decoherence. However, a critical observation is that as $\mu$ approaches $1$, these curves converge, indicating that strong classical correlations can mitigate the effects of decoherence, even for the infinite-time limit (see also Refs. \cite{Hu2019,Karpat2018,HaddadiLPL2022}).

Focusing specifically on Fig.  \ref{figure10} (c), we see that as $\mu$ approaches $1$, the Bell parameter converges to a limit close to $2\sqrt{2}$, which is the maximum possible value of non-locality. This is significant because it suggests that even after infinite time, with sufficient classical correlations, quantum non-locality can be preserved, indicating a resistance to complete degradation.

The physical interpretation of the results provided in Fig. \ref{figure10} indicates that when classical correlations are maximized ($\mu \rightarrow 1$), quantum resources are preserved. This robustness arises because classical correlations effectively synchronize the environment's effect on the system, reducing the differential impact of decoherence on quantum resources \cite{Hu2019}. The fact that these findings hold true even in the infinite-time limit highlights the importance of classical correlations in quantum information processing and suggests that maintaining high classical correlations can be a strategy to ensure long-term quantum resource preservation.

As we conclude this section, it is important to address further observations regarding the discussed phenomena. Firstly, the classical correlations between successive channel actions serve to enhance the overall coherence of both qubits collectively, without affecting the coherence of the single qubit. Secondly, we identify two potential sources of memory effects in the dephasing channel under scrutiny. The first originates from temporal correlations inherent in the time evolution of each qubit, while the second arises from classical correlations between successive channel actions \cite{Addis2016}. These factors collectively influence the decoherence process of the system. Notably, the damped oscillations of quantum resources are attributed to the first type of non-Markovian memory effects, while their increase at a fixed time is attributed to the second type of memory effects \cite{Hu2019}.

\section{Conclusion and outlook}\label{sec7}

In the realm of quantum mechanics and gravitational physics, the study of quantumness in gravitational cat states within correlated dephasing channels presents intriguing insights into the interplay between quantum coherence and classical correlations. Our investigation delved into the effects of classical correlations between successive actions of a dephasing channel on the decoherence dynamics of two gravitational cats, or two qubits, operating in a thermal regime.

We observed that the enhancement of classical correlations throughout the entire duration when the two qubits traverse the channel can substantially boost quantum coherence, local quantum Fisher information, and Bell non-locality. This enhancement suggests a nuanced relationship between classical and quantum correlations, demonstrating that classical information can play a constructive role in preserving and even amplifying quantum features under certain conditions.
However, the landscape of gravitational interaction and energy gap between states introduced complex influences on the quantum characteristics of gravitational cats. These intricate impacts hint at the multifaceted nature of gravitational interactions in quantum systems, where gravitational effects can either augment or hinder quantum coherence and correlations.

The discovery of these new features not only enriches our understanding of the fundamental principles governing gravitational cat states but also paves the way for advancements in quantum information processing. By harnessing classical correlations to enhance quantum coherence and information content, we may find novel avenues for improving the performance and reliability of quantum technologies.

Looking ahead, future research could delve deeper into the underlying mechanisms that drive the observed phenomena, exploring the specific conditions under which classical correlations exert the most significant influence on gravitational cat states. Additionally, investigations into the broader implications of these findings for gravitational physics, quantum computing, and quantum communication systems would be invaluable.

In summary, our study underscores the intricate interplay between classical correlations and quantum characteristics in gravitational systems, offering new perspectives and potential breakthroughs at the intersection of gravitational physics and quantum information science. As we continue to unravel the mysteries of quantum gravity, the insights gained from this research promise to shape the future landscape of both fields, opening doors to discoveries and innovative applications.

\vspace{20pt}

\section*{Acknowledgements}
This research is supported by the Postdoc grant of the Semnan University under Contract No. 21270.
\\
\\
\section*{Conflict of interest}
The authors declare that they have no known competing financial interests.
\\
\\
\section*{Data Availability Statement}
No datasets were generated or analyzed during the current study.

\end{document}